%% file: main.tex
\documentclass[10pt,twocolumn,letterpaper]{article}

\usepackage[pagenumbers]{cvpr} 

\input{preamble}

\def\paperID{7531} 
\def\confName{CVPR}
\def\confYear{2024}

\title{Kernel Diffusion: An Alternate Approach to Blind Deconvolution}

\author{Yash Sanghvi, Yiheng Chi, and Stanley H. Chan\\
School of Electrical and Computer Engineering, Purdue University\\ 
West Lafayette, IN \\
{\tt\small \{ysanghvi, chi14, stanchan\}@purdue.edu}
}

\begin{document}
\maketitle

\begin{abstract}
Blind deconvolution problems are severely ill-posed because neither the underlying signal nor the forward operator are not known exactly. Conventionally, these problems are solved by alternating between estimation of the image and kernel while keeping the other fixed. In this paper, we show that this framework is flawed because of its tendency to get trapped in local minima and, instead, suggest the use of a kernel estimation strategy with a non-blind solver. This framework is employed by a diffusion method which is trained to sample the blur kernel from the conditional distribution with guidance from a pre-trained non-blind solver. The proposed diffusion method leads to state-of-the-art results on both synthetic and real blur datasets. 
\end{abstract}

\section{Introduction}
\label{sec:intro}
Images taken by a hand-held camera under long exposure suffer from motion blur. Assuming this blur is spatially invariant, the captured image $\vy$ and the underlying sharp image $\vx$ are related as follows: 
\begin{align}
    \vy = \vk \circledast \vx + \vn ,
\end{align}
where $\vk$, $\vn$ refer to the blur kernel corresponding to the camera motion and noise, respectively, and the operator $\circledast$ is convolution. The inverse problem of estimating $\vx$ given $\vy$ is called deconvolution, which can be classified into \emph{non-blind} and \emph{blind}. 

In the non-blind case, the blur kernel $\vk$ is known. There are plenty of iterative methods which have been successfully applied to various image restoration tasks. Usually, these approaches, such as Plug-and-Play \cite{2013_PlugandPlay_SIP, 2016_P4IP_Elsevier, 2017_NonLinearPnP_SPL, 2019_OnlinePnP_TCI, 2016_PnPFixedPoint_TCI, 2020_PnPMRI_SPL} and Regularization by Denoising (RED) \cite{2017_RED_SIAMImaging, 2018_REDClarify_TCI, 2021_RED_PRO_SIAM}, use a pre-trained image denoiser as a substitute for a complex image prior. Recent deep learning counterparts to these approaches involve unrolling the iterative methods followed by end-to-end training \cite{2019_UnrollingDeblurring_ICASSP,2017_CNNPrior_CVPR,2016_ADMMNet_NeurIPS,2021_UnrollingReview_SPM}.

In blind deconvolution, the blur kernel $\vk$ is not known. A natural strategy for the blind inverse problem is to jointly estimate the blur kernel $\vk$, and the image $\vx$ by maximizing the posterior probability:
\begin{flalign}
    \begin{split}
    { \hat{\vx}, \hat{\vk}} &= \underset{\vx, \vk}{\text{arg}\max} \Big[ p(\vx, \vk | \vy)\Big] \\&=\underset{\vx, \vk}{\text{arg}\max} \Big[ p( \vy | \vx, \vk) p (\vx) p(\vk) \Big]  \\
    &= \underset{\vx, \vk}{\text{arg}\max} \Big[ \log p(\vy | \vk, \vx) + \log p(\vx) + \log p(\vk)\Big] .
    \end{split} \label{eq:joint_map_x_k}
\end{flalign}
The Maximum-A-Posteriori (MAP) for the pair $(\vx, \vk)$, i.e. $\text{MAP}_{\vx, \vk}$, can be obtained by an iterative scheme which alternates between the following minimization steps:
\begin{align}
    \vx^{i+1} &= \text{arg}\min\limits_{\vx} \Big[ \|\vy - \vx \circledast \vk^{i}\|^2  + \phi(\vx) \Big] , \\
    \vk^{i+1} &= \text{arg}\min\limits_{\vk} \Big[ \|\vy - \vx^{i+1} \circledast \vk\|^2  + \psi(\vk) \Big] ,
\end{align}
where $\phi(\cdot)$, $\psi(\cdot)$ represent the image and kernel priors. Many classical deconvolution techniques can be framed as variations of this framework \cite{1998_TVBlind_TIP,2013_L0Sparse_CVPR,2019_L0Gradient_IPOL,2008_HighQualityDeblur_TOG}. Even neural network based iterative methods such as Self-Deblur \cite{2020_SelfDeblur_CVPR} and diffusion models \cite{2023_BlindDPS_CVPR, 2023_GibbsDDRM_Arxiv} have achieved state-of-the-art performance in blind deconvolution using the stated framework.

Despite recent success of diffusion methods, they are based on $\text{MAP}_{\vx, \vk}$ approach, which was first brought into question by Levin et al. \cite{2011_LevinUnderstanding_PAMI}  They argued that the number of variables to be estimated, i.e., $\{\vx, \vk\}$, are much larger than the number of known variables, i.e., $\vy$, and as a result, \eqref{eq:joint_map_x_k} favors the no-blur solution ${\hat{\vx}, \hat{\vk}} = (\vy, \mI)$. \cite{2011_LevinUnderstanding_PAMI} proposed an alternate strategy of solving the kernel$\vk$ first to exploit the inherent symmetry in the problem. This is achieved by marginalizing out the image space as follows:
\begin{align}
    \hat{\vk} =  \underset{\vk}{\text{arg}\max} \Big[ p(\vk|\vy) \Big] = \underset{\vk}{\text{arg}\max} \Big[ \int p(\vk, \vx|\vy) d\vx\Big] .\label{eq:levin_kernel}  
\end{align}

\begin{figure*}[h]
    \centering
    \begin{tabular}{cccc}
    \includegraphics[width=0.24\linewidth]{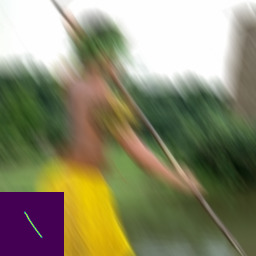} & 
    \hspace{-2.0ex}\includegraphics[width=0.24\linewidth]{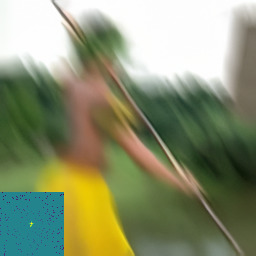} &
    \hspace{-2.0ex}\includegraphics[width=0.24\linewidth]{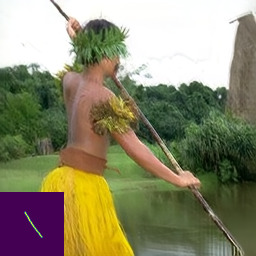} & 
    \hspace{-2.0ex}\includegraphics[width=0.24\linewidth]{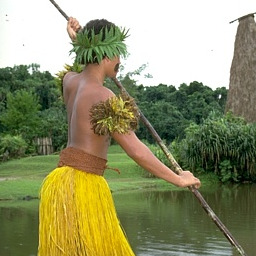}
     \\
     \makecell[t]{\small{(a) Blurred image } \\ \small{Inset: true kernel }} &
     \hspace{-2.0ex}\makecell[t]{\small{(b) Blind-DPS \cite{2023_BlindDPS_CVPR}, based on}\\ \small{alternating minimization} \\ \small{Inset: estimated kernel}} & 
     \hspace{-2.0ex}\makecell[t]{\small{(c) Kernel-Diff \textbf{(ours)}, based on} \\ \small{kernel estimation strategy} \\ \small{Inset: estimated kernel}} & 
     \hspace{-2.0ex}\makecell[t]{\small{(d) Ground Truth}}
    \end{tabular}

    \caption{In this paper, we present a novel diffusion-based approach to the blind deconvolution problem. The conventional approaches, such as Blind-DPS, iterate between between kernel and image space minimization. Our approach uses a kernel-first estimation strategy in the context of diffusion models to sample kernels from the conditional distribution $p(\vk|\vy)$.}
    \label{fig:introduction}
\end{figure*}

Although $\text{MAP}_{\vk}$ optimization can lead to significantly improved reconstruction (as shown in \fref{fig:introduction}(c) ) it has not been adopted in blind deconvolution solutions, which we surmise because of the intrinsic computational difficulty of the integration in \eqref{eq:levin_kernel}.

\subsection{Contributions}
In this paper, we address the problem with a kernel estimation approach, and we demonstrate that this strategy can be implemented in practice. Our contribution in this paper can be summarized as follows:
\begin{enumerate}
    \item Using a simple 1D deconvolution example, we demonstrate how the $\text{MAP}_{\vx,\vk}$ framework can be susceptible to local minima while the suggested strategy of $\text{MAP}_{\vk}$ does not. Using this example, we show how the marginalization in \eqref{eq:levin_kernel} can be achieved in practice using a non-blind solver. 
    \item Next, we propose a novel diffusion method, called Kernel-Diffusion (\emph{Kernel-Diff}) which is based on the design principles outline dabove and is trained to sample from the conditional distribution $p(\vk|\vy)$. This diffusion method coupled with a differentiable non-blind solver achieves state-of-the-art deconvolution performance. We show with numerical experiments on real and synthetic data that the proposed approach creates significant performance improvement. 
\end{enumerate}

\section{Background}
\label{sec:related}
\subsection{Diffusion Methods}
Diffusion models \cite{2015_Diffusion_ICML, 2020_DDPM_NeurIPS, 2011_DenoisingScoreMatching_NC, 2021_ScoreBasedDiffusion_ICLR} are a popular generative method which can be used to sample any target diffusion $p_{data}(\vx)$. The diffusion models begin with a forward process, which starts from $\vx_0 \sim p_{data}(\vx)$ and gradually adds noise according to variance schedule $\{ \beta_1, \beta_2, ... \beta_T \}$, formulated as
\begin{align}
    \vx_t \sim  \mathcal{N}(\sqrt{1-\beta_t}\vx_{t-1}, \beta_t \mI)
\end{align}
at time steps $t \in \{1, 2, \dots, T\}$. This forward process converges to $\vx_T \sim \mathcal{N}(0, \mI)$. Next, the reverse diffusion process is used to sample from the distribution of $p_{data}$ by starting with $\vx_T \sim \mathcal{N}(0, \mI)$ and performing the iterative updates
\begin{align}
    \begin{split}
    \vx_{t-1} = \frac{1}{\sqrt{\alpha_t}}\Big(\vx_t - \frac{1-\alpha_t}{\sqrt{1-\bar{\alpha_t}}}\epsilon_{\theta^*}(\vx_t, t)\Big) + \sigma_t \vz_t , \\ \vz_t \sim \mathcal{N}(0, \mI) ,
    \end{split}
    \label{eq:reverse_diffusion}
\end{align}
where $\alpha_t \bydef 1-\beta_t$, $\bar{\alpha_t} \bydef \prod\limits_{i=1}^t \alpha_i$, and $\epsilon_{\theta^*}(\vx_t, t)$ is the trained diffusion network used to learn to denoise $\vx_t$ at any given time step, trained by minimizing the following loss function.
\begin{align}
    \begin{split}
    \mathcal{L}(\theta) \bydef \E_{t, \vx_0, \epsilon}\Big[ \| \epsilon - \epsilon_{\theta}(\sqrt{\bar{\alpha}_t}\vx_0    + \sqrt{1-\bar{\alpha_t}}\epsilon,t) \|^2 \Big] .
    \end{split}
\end{align}

\textbf{Inverse Problems and Diffusion Models}.
Score based diffusion methods have been extensively used in solving inverse problems \cite{2022_DDRM_NeurIPS, 2023_3DInverse_CVPR, 2023_DDPnP_CVPRW, 2021_SNIPS_NeurIPS, 2023_Soft_TMLR, 2021_StochasticDenoising_CVPR, 2021_StochasticLinearInverseProblems_NeurIPS} by sampling from the posterior $p(\vx | \vy)$ and the corresponding score is calculated using the Bayes' formula
\begin{align}
    \nabla_{\vx_t} \log p(\vx_t|\vy) = \nabla_{\vx_t} \log p(\vx_t) + \nabla_{\vx_t} \log p(\vy|\vx_t) . \label{eq:posterior_score}
\end{align}
For example, \cite{2021_MRIDiffusion_NeurIPS} uses an approximation to the posterior score in \eqref{eq:posterior_score} and used the Langevin dynamics to sample from the distribution $p(\vx|\vy)$ for the MRI compressed sensing problem; \cite{2022_DiffuseFaster_CVPR,2022_MCGDiffusion_NeurIPS} solve the inpainting and super-resolution inverse problems with a modified reverse diffusion scheme which includes a projection step to reduce the consistency metric $\|\vy- \mA\vx\|$ in each iteration; diffusion Posterior Sampling (DPS) \cite{2023_DPS_ICLR} solves the noisy inverse problem with alternating diffusion and gradient descent steps. 

\textbf{Image Restoration and Diffusion Models} Diffusion models sample from a target distribution while MMSE estimators average over them \cite{2021_StochasticDenoising_CVPR}. As a result, the former has been widely used in image restoration tasks such as deblurring \cite{2022_StochasticRefinement_CVPR}, super-resolution \cite{2021_DiffusionSR_arxiv, 2022_SRDiff_Neurocomputing}, and low-light enhancement \cite{2023_LowLightDiff_ICCV,2023_LowLightWavketDiff_Arxiv}. For an exhaustive review, we refer the reader to \cite{2023_DiffIRSurvey_Arxiv}.

\textbf{Guidance in Diffusion Models}. Pre-trained diffusion models are often guided at each iteration to produce desired results. This is first proposed in \cite{2021_DifffusionUNet_NeurIPS} where a classifier gradient is added to each iteration for the final output to belong to the given class. ILVR \cite{2021_ILVR_ICCV} guides the image diffusion process by constraining the low-frequency information of an image. In the same spirit, \cite{2022_ZeroShotGuidance_ICLR} guides the diffusion process by refining only the null-space of the corresponding inverse problem. \cite{2022_StochasticRefinement_CVPR} guides a deblurring diffusion model using a pretrained structure module.

\subsection{Blind Deconvolution}
Traditional blind deconvolution methods can be categorized under the $\text{MAP}_{\vx,\vk}$ framework \cite{1998_TVBlind_TIP,2013_L0Sparse_CVPR,2019_L0Gradient_IPOL,2008_HighQualityDeblur_TOG, 2016_DarkChannel_CVPR, 2020_Efficient_TCI, 2017_ExtremeChannel_CVPR, 2013_EdgeBlur_ICCP, 2019_LocalGradient_CVPR}. For example, \cite{2009_FastDeblurring_TOG} uses salient edges for the kernel estimation and then performs image estimation using a Total Variation (TV) prior.  \cite{2010_TwoPhase_ECCV} refines the kernel estimation process by iterative support detection.  \cite{2010_DeblurringMDF_ECCV} performs alternating image and kernel minimization but uses a motion density representation for the blur kernel instead of for pixels. Self-Deblur \cite{2020_SelfDeblur_CVPR} uses an alternating minimization scheme where a latent deep image prior \cite{2018_DIP_CVPR} is used to represent the kernel and image space. 

Deep learning methods have replaced the optimization-based approaches and usually involve end-to-end training of image restoration networks \cite{2015_LearnToDeblur_PAMI, 2015_LearnToDeblurCNN_CVPR, 2018_SRN_CVPR, 2018_DeblurGAN_CVPR, 2019_DHMPN_CVPR, 2021_MPRNet_CVPR, 2021_UFormer_CVPR,  2022_Restormer_CVPR} using large datasets of pairs of sharp and blurry images, e.g., GoPro \cite{2017_GoPro_CVPR} and RealBlur \cite{2020_RealBlur_ECCV} datasets. For a more exhaustive review of deep learning based approaches, we refer the readers to \cite{2022_ReviewDeblurring_IJCV}.

A few methods have utilized the $\text{MAP}_{\vk}$ approach, even though not always explicitly stated by the authors. \cite{2016_NeuralDeblurring_ECCV} uses a fully convolutional network to predict the Fourier coefficients of the underlying kernel which are used to deconvolve the blurry image in Fourier space. \cite{2023_PixelWiseKernel_TCI, 2021_NUBlurKernel_Arxiv} train a Kernel-Prediction Network for all pixels in the image to deal with spatially varying blur. \cite{2022_IterativeKernel_TCI} leverages the differentiablity of a non-blind solver \cite{2022_NonBlindUnrolling_TCI} to frame the blind deconvolution problem entirely in terms of kernel estimation problems. This strategy is improved by \cite{2023_StructuredKernel_CVPR} using a key-points based representation to further reduce the dimensionality of the problem. 

\begin{figure*}
    \centering
    \begin{tabular}{cc}
        \hspace{-5.0ex}\includegraphics[trim={0 60 0  0},clip,width=0.99\linewidth]{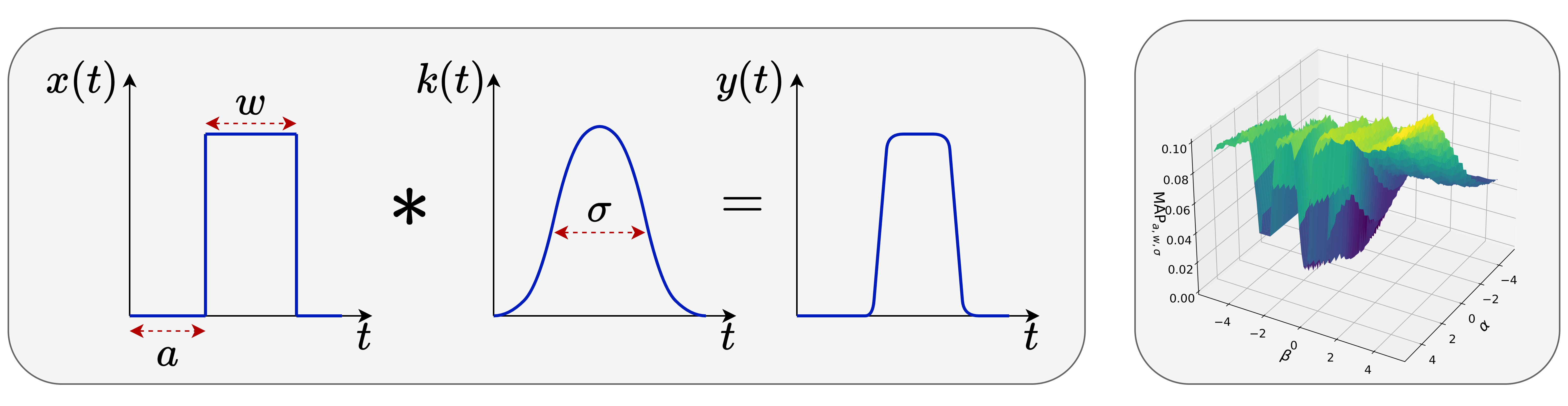}   
        &  \\
        \hspace{-35.0ex}\makecell{\small{\textbf{Toy 1D Deconvolution}: Estimate image space $\{a, w\}$} \small{and blur $\sigma$ from $\vy(t)$}} & \hspace{-32.0ex}\makecell{\small{Loss surface -}  $\text{MAP}_{a,w,\sigma}$} \\
        \hspace{-6.0ex}\includegraphics[trim={0 150 0 0},clip,width=0.9\linewidth]{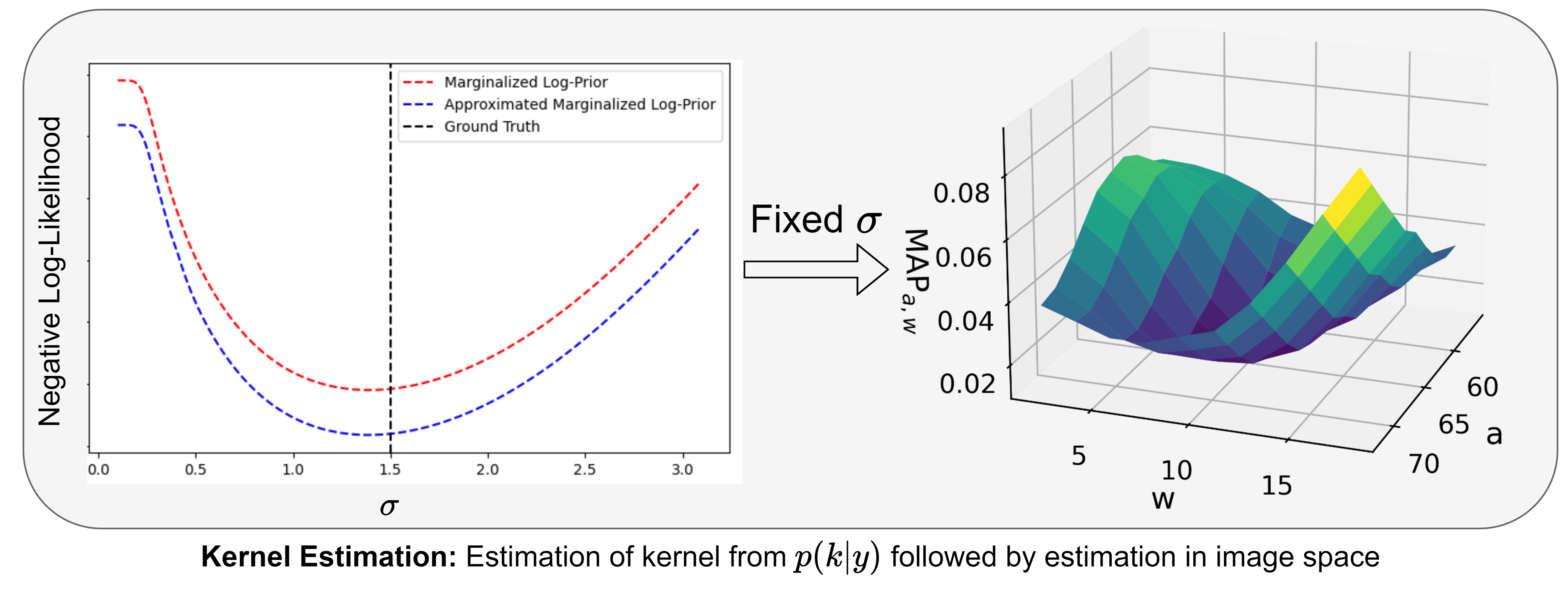} & \\
        \makecell{\small{\textbf{Kernel Estimation} from $p(\vk|\vy)$ followed by image estimation}}
    \end{tabular}
    
    \caption{\textbf{Perils of Alternating Minimization}: (Top-Left) Example blind deconvolution setup. Given the blurred signal $y(t)$, we need to recover the underlying latent signal $x(t)$, characterized by $\{a, w\}$, blurred by Gaussian kernel of width $\sigma$ (Top-Right) Loss function for conventional image-kernel minimization, projected in 2D. Alternating minimization of image and kernel can be easily trapped in multiple local minima leading to convergence to false solutions. (Bottom) $\log p(\sigma | y)$ and our suggested approximation from \eqref{eq:approximation}. The suggested strategy is to first estimate the kernel by estimating the maximum from $p(\vk|\vy)$ followed by estimation of the image space parameters using the estimated kernel. In this case, we get two convex surfaces and hence the overall scheme is less likely to get trapped in the local minima.}
    \label{fig:toy-example}
\end{figure*}

\section{Method}
\label{sec:method}

\subsection{Deconvolution Scheme Analysis}
\label{subsec:motivation}
We motivate the proposed method by analyzing a simple 1D deconvolution problem to demonstrate the inherent difficulties of optimization in the alternating minimization framework. We show that minimizing the cost function in \eqref{eq:joint_map_x_k} is very likely to converge to false minima. We also show the benefits of the $\text{MAP}_{\vk}$ strategy.

\textbf{Simple deconvolution problem}. Assume an unknown pulse function $\vx_{a, w}(t)$ where $a$ represents the beginning of the pulse and $w$ represents its width. The signal is blurred by a Gaussian pulse $\vk_{\sigma}(t)$ where $\sigma$ represents its standard deviation. The resulting signal $\vy(t)$ is obtained by blurring the signal $\vx(t)$ followed by additive Gaussian noise, i.e.,
\begin{align}
    \vy =  \vk_{\sigma} \circledast \vx_{a,w} + \mathcal{N}(0, \beta \mI) .
\end{align}
Note that that in this setup the image $\vx$ and kernel $\vk$ are characterized completely be scalars $\{a, w\}$ and $\sigma$, respectively. Therefore, the blind deconvolution problem can be considered as an estimation of the three variables $\{a, w, \sigma\}$. 

\textbf{Alternating Methods versus Kernel Minimization}.
For the stated example, the conventional alternating minimization approach of $\text{MAP}_{\vx, \vk}$ can be formulated as
\begin{flalign}
\begin{split}
    \hat{a}, \hat{w}, \hat{\sigma} &= \underset{a, w, \sigma \geq 0}{\text{arg}\max} \Big[p(a, w, \sigma | \vy)\Big]  \\ 
    &= \underset{a, w, \sigma \geq 0}{\text{arg}\max} \Big[ \exp\big( -\frac{1}{2\beta}\| \vy - \vx_{a,w}\circledast\vk_{\sigma} \|^2\big) \Big].
\end{split}
    \label{eq:image_and_kernel_minimization}
\end{flalign}
The equivalent $\text{MAP}_{\vk}$ would be first estimating the kernel width $\sigma$ as
\begin{align}
\begin{split}
    \hat{\sigma} &= \underset{\sigma \geq 0}{\text{arg}\max} \Big[ p(\sigma | \vy)\Big] \\
    &= \underset{\sigma \geq 0}{\text{arg}\max} \Big[ \int p(a, w, \sigma | \vy)da\; dw\Big]   \\  &=\underset{\sigma \geq 0}{\text{arg}\min}\Big[\sum\limits_{a,w\geq 0}\exp\big[ -\frac{1}{2\beta}\|\vy-\vx_{a,w}\circledast\vk_{\sigma}\|^2 \big]\Big],
\end{split}\label{eq:kernel_minimization}
\end{align}
followed by the estimation of $\{a, w\}$:
\begin{align}
    \{ \hat{a}, \hat{w} \}  &= \underset{ a, w \geq 0}{\text{arg}\min} \Big[\|\vy-\vk_{\hat{\sigma}}\circledast\vx_{a,w}\|^2 \big]\Big].
\end{align}

\textbf{Analysis}. In Figure \ref{fig:toy-example}, we plot the loss functions corresponding to the two schemes outlined in \eqref{eq:image_and_kernel_minimization} and \eqref{eq:kernel_minimization}. We have the following observations according to the loss functions. First, for the joint alternating minimization scheme, the loss function contains many local minima. Even though the global minimum is close to the ground-truth, it is very difficult for a gradient-descent based solver to converge to it unless the random initialization is already very close. Second, for the kernel-based minimization scheme which reduces the search space of the optimization to kernel only, the corresponding loss function has only one minimum, which is also close to the ground-truth. It is very likely that the estimated $\vx$, i.e., $\{a,w\}$, will converge to this minimum. Therefore, the kernel-based minimization scheme can consistently produce significantly better results with much lower prediction error than the alternating scheme.

\begin{figure*}
    \centering
    \includegraphics[trim={0 0 0 0},clip,width=0.95\linewidth]{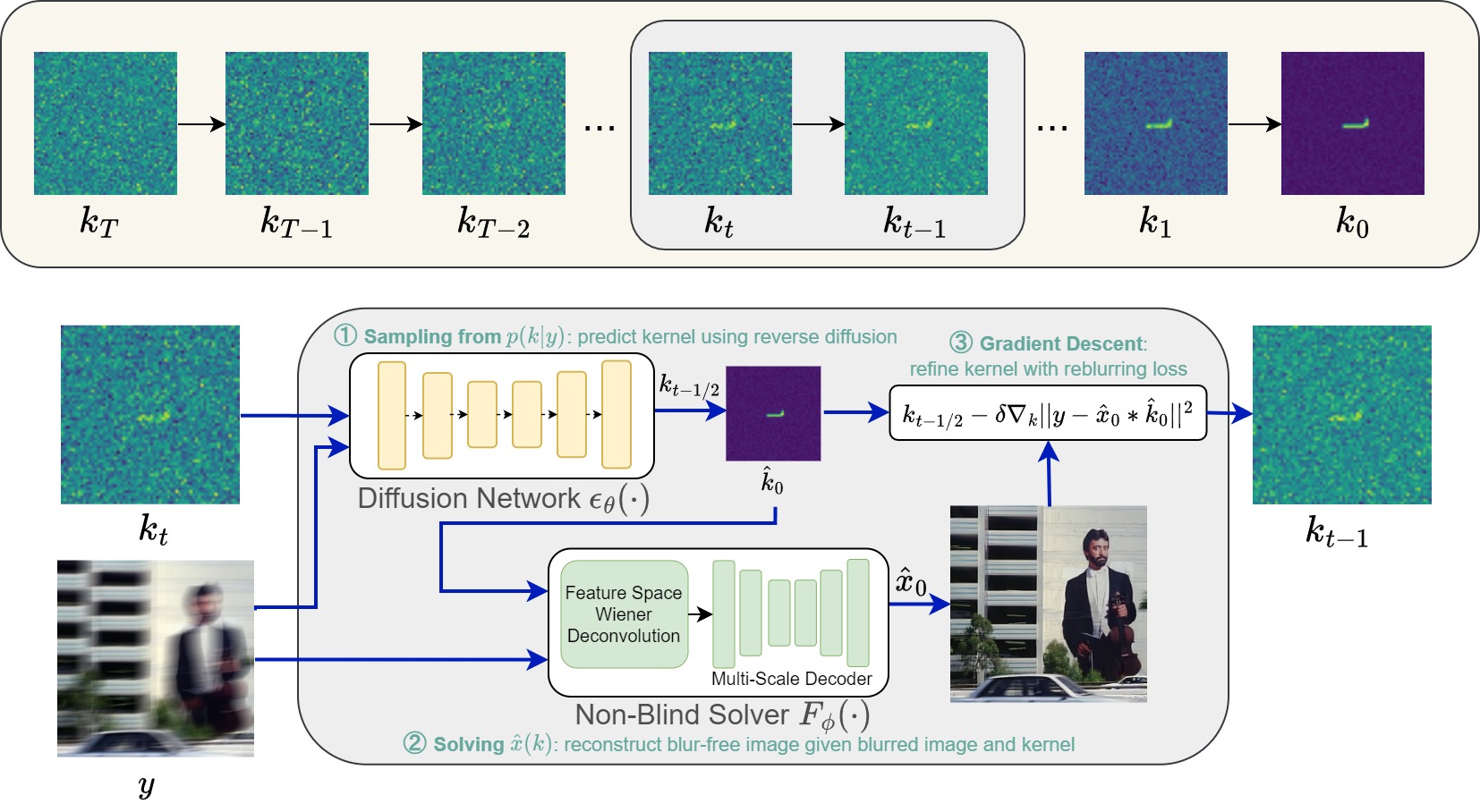}   
    \caption{\textbf{Proposed Kernel-Diff}. Each iteration of the proposed kernel-based diffusion contains two sub-steps. First, we perform a reverse diffusion step using the trained diffusion model $\epsilon_{\theta}(\vk_t, \vy, t)$. Next we plug the temporary kernel estimate $\hat{\vk}_0$ in the differentiable non-blind solver $F(\cdot)$ and perform a gradient descent update  to guide the diffusion process to minimizing the reblurring loss $\|\vy- \hat{\vk}_0 \circledast \hat{\vx}_0\|^2$.}
    \vspace{-2ex}
    \label{fig:diffusion_proposed}
\end{figure*}

\subsection{Marginalization of Image Space}
In practice, the estimation of $p(\vk | \vy)$ can be difficult because the marginalization in \eqref{eq:levin_kernel} requires a weighted average over all images. We surmise that due to lack of a feasible method to evaluate the likelihood, the approach from \cite{2011_LevinUnderstanding_PAMI} was not explored by the imaging community for the blind deconvolution. To mitigate the computational issue, we propose an approximation to the marginalization using a non-blind solver. 

Before stating the approximation, we explicitly define a non-blind solver. Given a blurred image $\vy$ and the corresponding blur kernel $\vk$, we can easily estimate the latent image using a non-blind solver which performs the following image estimation:
\begin{align}
    \hat{\vx}(\vk) = \text{arg}\max\limits_{\vx} \Big[ p(\vx| \vk , \vy) \Big] \bydef  F(\vy, \vk). \label{eq:best_image_based_on_kernel}
\end{align}
Traditionally, these non-blind solvers are implemented using an iterative method where the corresponding cost function for MAP estimation in \eqref{eq:best_image_based_on_kernel} is minimized using gradient descent or proximal updates. Convolutional neural networks with Wiener-deconvolution processing have been used with great success also \cite{2020_DeepWiener_NIPS, 2022_NonBlindUnrolling_TCI, 2018_LearningDataTerms_ECCV, 2020_ModelUncertrainity_CVPR, 20203_DeepRL_IJCV}. The latter provides us with $F(\cdot)$ as a differentiable function with respect to $\vk$. The significance of this property will become obvious to the reader in the next subsection. 

Now, given our definition of a non-blind solver, \emph{we approximate the conditional probability $p(\vk | \vy)$ as follows}:
\begin{align}
\begin{split}
    p(\vk | \vy ) = \int p(\vx, \vk | \vy) d\vx \approx  \frac{1}{Z}p(\hat{\vx}(\vk), \vk | \vy) ,
\end{split}
     \label{eq:approximation}
\end{align}
where the integration is performed using the Laplace approximation around the mode of the conditional distribution $p(\vx, \vk |\vy)$ and $Z$ is the constant scaling factor. For a more detailed explanation of this approximation, we refer the readers to Section {\color{red}C} of the supplementary document.

For the problem in Section \ref{subsec:motivation}, $p(\sigma|\vy)$ can be approximated as
\begin{align}
    p(\sigma | \vy) \approx \frac{1}{Z'}\exp\Big[ -\frac{1}{2\beta^2} \|\vy - \vx_{a(\sigma), w(\sigma)}\circledast \vk_{\sigma}\|^2\Big] ,
\end{align}
where $\{a (\sigma), w(\sigma)\}$ represents the best estimate of the image space parameters given $\sigma$ according to \eqref{eq:best_image_based_on_kernel} and $Z'$ is a constant scalar.

Through the approximation in \eqref{eq:approximation}, we can now estimate the kernel by maximizing $p(\vk|\vy)$. Firstly, \eqref{eq:approximation} removes the intractable marginalization of the image space making to easier to evaluate the likelihood. Secondly, using the non-blind solver i.e. $\hat{\vx}(\vk)$ allows us to convert the an image-kernel optimization problem to an purely kernel-estimation problem. 

\subsection{Kernel-Based Diffusion}
\label{subsec:diffusion-method}

To sample from $p(\vk|\vy)$, we propose to use a kernel-based diffusion network. This is different from existing methods using diffusion models, e.g., Blind-DPS~\cite{2023_BlindDPS_CVPR}, that resemble the alternating minimization approach, aiming to sample $\{\vx, \vk\}$ from the conditional distribution $p(\vx, \vk | \vy)$ using implicit score models for image and kernel spaces.

\textbf{Diffusion Network}. 
We use a modified version of the U-Net architecture as opposed to the one described in \cite{2021_DifffusionUNet_NeurIPS}. Since we are sampling from a conditional distribution, the diffusion model $\epsilon_\theta(\vk, \vy, t)$ takes as input both the current kernel estimate $\vk_{t}$ and the blurred image $\vy$. This is achieved in a similar fashion as \cite{2021_DiffusionSR_arxiv}. Specifically, our U-Net architecture has two encoders, one for the image $\vy$ and the other for the kernel $\vk$. In our implementation, we use images of size $256 \times 256$ and kernels of size $64 \times 64$. At the end of the encoders, the encoded features of the kernel branch are upscaled by a factor of 4 and concatenated to the encoded features of the image branch. The concatenated features are then passed through a common U-Net decoder stage to predict the kernel at the next time step.

The diffusion model $\epsilon_{\theta}(\cdot)$ is trained by minimizing the following loss function:
\begin{align}
    \begin{split}
    \mathcal{L}_{\text{kernel}}(\theta) &\bydef \E_{t, \vk, \vy, \epsilon}\Big[ \| \epsilon - \epsilon_{\theta}(\sqrt{\bar{\alpha}_t}\vk  + \sqrt{1-\bar{\alpha_t}}\epsilon, \vy, t) \|_2^2 \Big] 
    \end{split}
\end{align}
where the expectation is taken over all timesteps $t$, blurred images $\vy$ and corresponding blur kernels $\vk$. With trained parameters $\theta^*$, we sample $p(\vk|\vy)$ using a series of reverse diffusion steps starting from $\vk_T \sim \mathcal{N}(0, \mI)$ and the following reverse-diffusion iteration:
\begin{align}
   \vk_{t-1} = \frac{1}{\sqrt{\alpha_t}}\bigg(\vk_{t} - \frac{1-\alpha_t}{\sqrt{1-\bar{\alpha_t}}}\epsilon_{\theta^*}(\vk_t, \vy, t)\bigg) + \bar{\sigma}_t \vz_t \label{eq:reverse} 
\end{align}

\textbf{Guidance using Non-Blind Solver}. Once we have sampled the kernel $\vk_0$, we estimate the latent image as $\vx_0 = F_{\phi}(\vy, \vk_0)$, where $F_{\phi}(\cdot)$ is a neural network non-blind solver. There is a vast selection of candidate non-blind solver, including many network-based methods proposed recently \cite{2020_DeepWiener_NIPS, 2020_LCHQS_ECCV, 2020_RGDN_TNN, 2021_SpatiallyVariant_CVPR, 2022_NonBlindUnrolling_TCI, 2023_INFWIDE_TIP}. These networks usually convert the blurred images to features space, apply Wiener deconvolution using the kernel followed by converting the deconvolved features into a single image using a decoder.

While diffusion processes are effective in generating samples from the conditional distribution $p(\vk | \vy)$, they do not guarantee minimization of the reblurring loss, i.e., $\|\vy - \vk\circledast \vx\|^2$. \emph{To minimize reblurring loss while also generating a plausible sample $p(\vk |\vy)$, we guide the reverse diffusion process using the pre-trained non-blind solver as follows.} 

At step $t$ of the reverse diffusion process, the kernel estimate is given as $\vk_{t}$. Using the trained diffusion model $\epsilon_{\theta^*}(\vk_t, \vy, t)$ and $\vk_t$, we can compute a temporary estimate of the sampled kernel $\vk_0$ as follows:
\begin{align}
    \hat{\vk}_0 = \frac{1}{\sqrt{\bar{\alpha_t}} }\big(\vk_{t} - \sqrt{1-\bar{\alpha}_t}\epsilon_{\theta^*}(\vk_t, \vy, t)\big) .
\end{align}
We then estimate the image using this kernel estimate $\hat{\vk}_0$ by $\hat{\vx}_0 \bydef F(\vy, \hat{\vk}_{0})$, plug into the reblurring loss, and perform gradient descent by

\begin{align}
\vk_{t} \leftarrow \vk_t - \delta\nabla_{\vk_t}\|\vy -  \hat{\vk}_0 \circledast F(\vy,\hat{\vk}_0) \|^2 .
\end{align}

\begin{algorithm}[b]
\begin{algorithmic}[1]
\State \textbf{Input}: Blurred Image $\vy$, Diffusion UNet $\epsilon_{\theta}(.)$,  Non-Blind Solver $F_{\phi}(.)$.
\State \textit{// Stage I: Reverse Diffusion}
\State $\vk_T \sim \mathcal{N}(\mathbf{0}, \mI)$
    \For{$t = T, \cdot\cdot\cdot, 1$}
    \State $\vz_t \sim \mathcal{N}(\mathbf{0}, \mI)$
    \State $\vk_{t-1/2} = \frac{1}{\sqrt{\alpha_t}}\bigg(\vk_{t} - \frac{1-\alpha_t}{\sqrt{1-\bar{\alpha_t}}}\epsilon_{\theta}(\vk_t, \vy, t)\bigg) + \sigma_t \vz_t $
    \State $\hat{\vk}_0 = \frac{1}{\sqrt{\bar{\alpha_t}} }\big(\vk_{t} - \sqrt{1-\bar{\alpha}_t}\epsilon_{\theta}(\vk_t, \vy, t)\big)$
    \State $\hat{\vx}_0 = F_{\phi}(\vy, \hat{\vk}_0)$
    \State $\vk_{t-1} =  \vk_{t-1/2} - \delta\nabla_{\vk_t}(\|\vy - \hat{\vk}_0\circledast \hat{\vx}_0 \|^2)$
    \EndFor
\State \textit{// Stage II: Refinement}
\For{$j = 1, \cdot\cdot\cdot, J$}
    \State $\vk_0 \leftarrow  \vk_0 - \delta\nabla_{\vk_0}(\|\vy - \vk_0 \circledast F_{\phi}(\vy, \vk_0) \|^2)$  
    \EndFor 
\State return $\vk_0$ and $\vx_{0} = F_{\phi}(\vy,\vk_0)$
\end{algorithmic}
\caption{Kernel-Diff}
\label{alg:diffusion_with_kernel}
\end{algorithm}

\textbf{Overall procedure}. We summarize the overall procedure in Algorithm \ref{alg:diffusion_with_kernel} and pictorially represent it in \fref{fig:diffusion_proposed}. At any given iteration $t$, we first perform reverse diffusion process using the denoising network $\epsilon_{\theta}(\cdot)$ to output $\vk_{t-1/2}$ (Step 5). Next, we perform a gradient descent update on $\vk_{t-1/2}$ the reblurring loss $\|\vy - \hat{\vk}_0 \circledast \hat{\vx}_0\|^2$ (Step 7-9). For further performance boost, once the reverse diffusion process is over, we repeat the gradient descent steps (Step 13) to refine the kernel, similar to Stage II in \cite{2023_StructuredKernel_CVPR}.

\section{Experiments}
\label{sec:experiments}

\subsection{Experiment Setup}
\textbf{Implementation of Kernel-Diff}. The diffusion network we use is a modified U-Net \cite{2021_DifffusionUNet_NeurIPS}. To train the diffusion model $\epsilon_{\theta}(\vk_t, \vy, t)$, we generate blurred data using synthetic motion blur kernels, with details of data synthesis to be elaborated in the next section. We assume symmetric boundary conditions during the convolution process. We train the diffusion U-Net architecture for 500,000 iterations with a batch size of 16. For the non-blind solver, we use the pre-trained network Deep Wiener Deconvolution (DWDN) \cite{2020_DeepWiener_NIPS} provided by the authors and use it in our scheme without any fine-tuning. 

\textbf{Training Dataset}. The diffusion U-Net in \emph{Kernel-Diff} as well as all methods we compare in our experiments are trained using the same set of blurred and clean image data. We synthesize the blurred images by applying random blur kernels from a kernel database on clean images, similar to \cite{2023_BlindDPS_CVPR}. The kernel database consists of 60,000 $64 \times 64$ blur kernels generated using an open-sourced motion blur model~\cite{LeviBorodenkoMotionBlur_Github}. For clean images, we use the full-resolution images of Flickr2K \cite{Flickr2K}. To train the diffusion U-Net for \emph{Kernel-Diff}, we additionally randomly crop the images to $128 \times 128$ patches and augment the data by random horizontal flipping.

\begin{figure*}[h]
    \begin{tabular}{ccccccc}
        \includegraphics[width=0.14\linewidth]{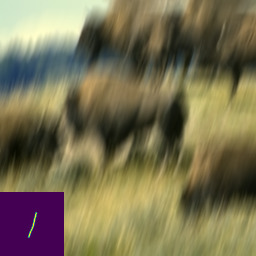} &
        \hspace{-2.0ex}\includegraphics[width=0.14\linewidth]{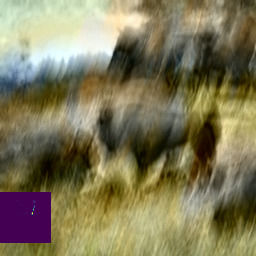} &
        \hspace{-2.0ex}\includegraphics[width=0.14\linewidth]{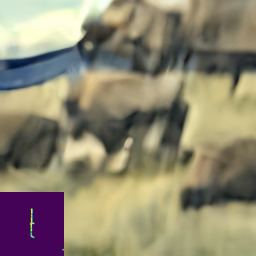} &
        \hspace{-2.0ex}\includegraphics[width=0.14\linewidth]{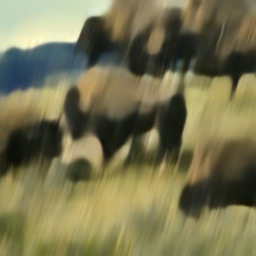} & 
        \hspace{-2.0ex}\includegraphics[width=0.14\linewidth]{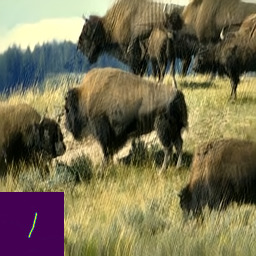} &
        \hspace{-2.0ex}\includegraphics[width=0.14\linewidth]{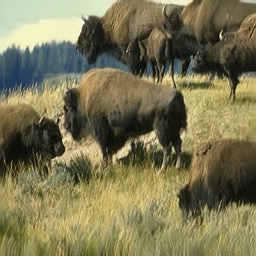} &
        \hspace{-2.0ex}\includegraphics[width=0.14\linewidth]{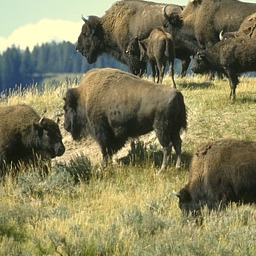}\\

        \includegraphics[width=0.14\linewidth]{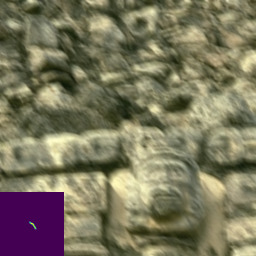} &
        \hspace{-2.0ex}\includegraphics[width=0.14\linewidth]{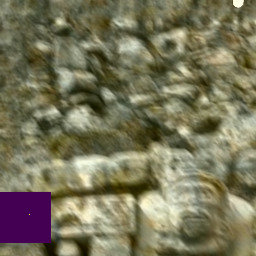} &
        \hspace{-2.0ex}\includegraphics[width=0.14\linewidth]{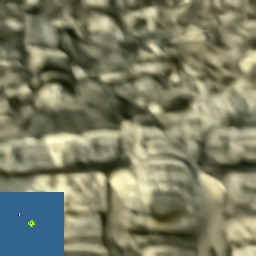} &
        \hspace{-2.0ex}\includegraphics[width=0.14\linewidth]{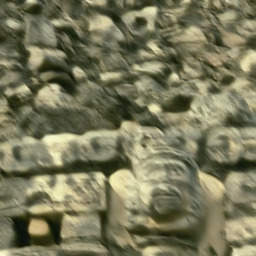} & 
        \hspace{-2.0ex}\includegraphics[width=0.14\linewidth]{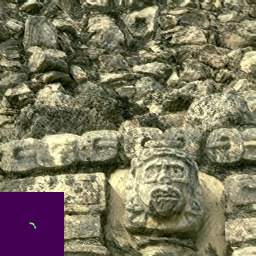} &
        \hspace{-2.0ex}\includegraphics[width=0.14\linewidth]{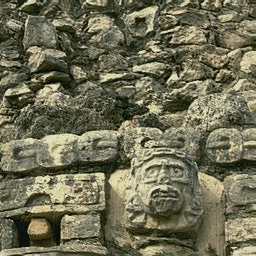} &
        \hspace{-2.0ex}\includegraphics[width=0.14\linewidth]{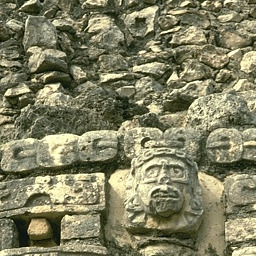}\\

        \includegraphics[width=0.14\linewidth]{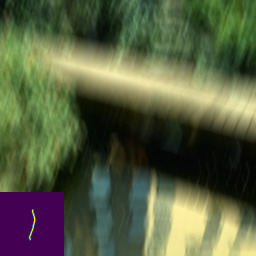} &
        \hspace{-2.0ex}\includegraphics[width=0.14\linewidth]{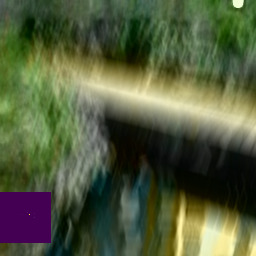} &
        \hspace{-2.0ex}\includegraphics[width=0.14\linewidth]{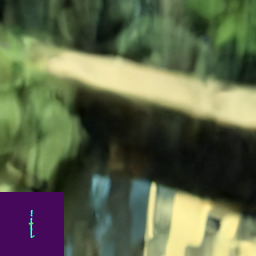} &
        \hspace{-2.0ex}\includegraphics[width=0.14\linewidth]{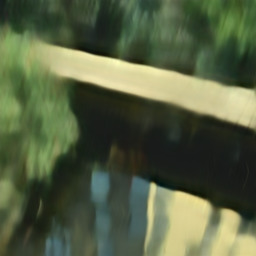} & 
        \hspace{-2.0ex}\includegraphics[width=0.14\linewidth]{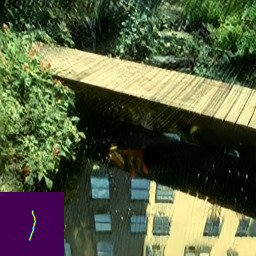} &
        \hspace{-2.0ex}\includegraphics[width=0.14\linewidth]{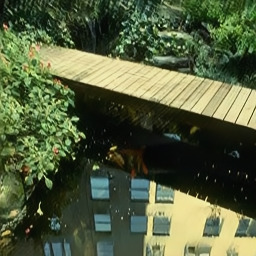} &
        \hspace{-2.0ex}\includegraphics[width=0.14\linewidth]{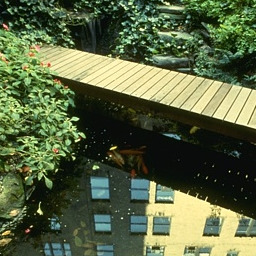}\\
        
        \includegraphics[width=0.14\linewidth]{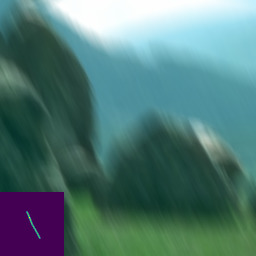} &
        \hspace{-2.0ex}\includegraphics[width=0.14\linewidth]{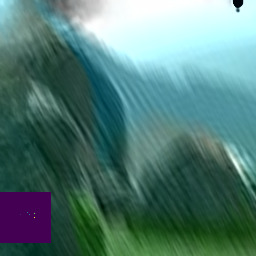} &
        \hspace{-2.0ex}\includegraphics[width=0.14\linewidth]{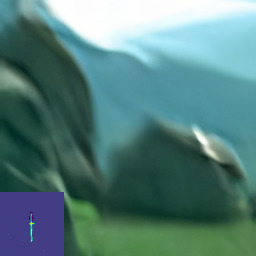} &
        \hspace{-2.0ex}\includegraphics[width=0.14\linewidth]{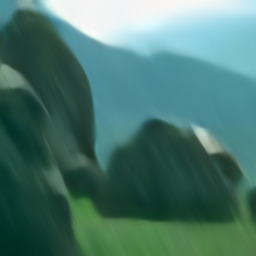} & 
        \hspace{-2.0ex}\includegraphics[width=0.14\linewidth]{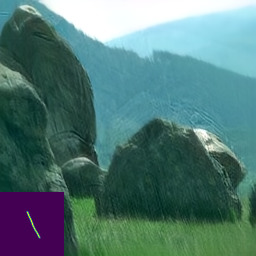} &
        \hspace{-2.0ex}\includegraphics[width=0.14\linewidth]{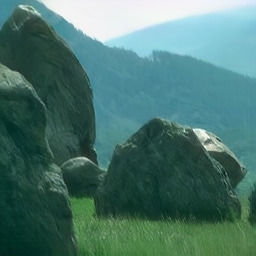} &
        \hspace{-2.0ex}\includegraphics[width=0.14\linewidth]{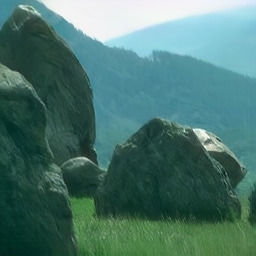}\\

        \includegraphics[width=0.14\linewidth]{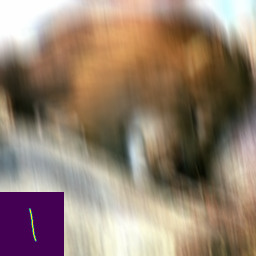} &
        \hspace{-2.0ex}\includegraphics[width=0.14\linewidth]{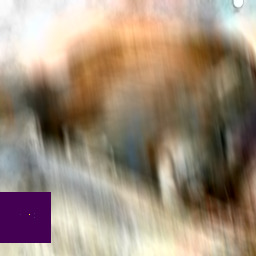} &
        \hspace{-2.0ex}\includegraphics[width=0.14\linewidth]{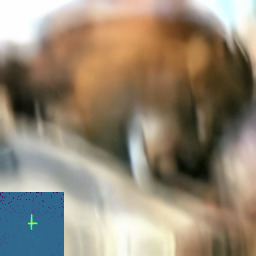} &
        \hspace{-2.0ex}\includegraphics[width=0.14\linewidth]{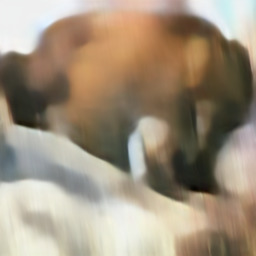} & 
        \hspace{-2.0ex}\includegraphics[width=0.14\linewidth]{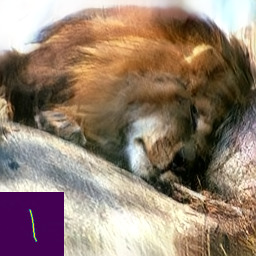} &
        \hspace{-2.0ex}\includegraphics[width=0.14\linewidth]{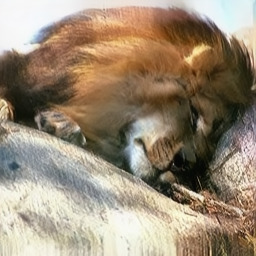} &
        \hspace{-2.0ex}\includegraphics[width=0.14\linewidth]{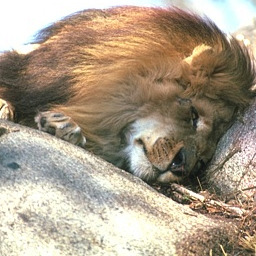} \\
        \makecell{\small{Blurred image} \\ \small{Inset: true kernel} } & 
        \hspace{-2.0ex}\makecell{\small{Self-Deblur} \\ \small{\cite{2020_SelfDeblur_CVPR}}} &
        \hspace{-2.0ex}\makecell{\small{Blind-DPS} \\ \small{\cite{2023_BlindDPS_CVPR}}} & 
        \hspace{-2.0ex}\makecell{\small{MPR-Net} \\ \small{\cite{2021_MPRNet_CVPR}}} & 
        \hspace{-2.0ex}\makecell{\small{\textbf{Kernel-Diff}} \\ \small{\textbf{(Ours)}}} & 
        \hspace{-2.0ex}\makecell{\small{DWDN-Oracle\textsuperscript{\red{\dag}}} \\  \small{\cite{2020_DeepWiener_NIPS}}} & 
        \hspace{-2.0ex}\makecell{\small{Ground Truth}} \\
    \end{tabular}
    \caption{\textbf{Qualitative Results}: Reconstruction results on synthetically blurred images from the BSD100 dataset. For Self-Deblur, Blind-DPS and Kernel-Diff, we show the estimated kernels in inset. \textsuperscript{\red{\dag}}DWDN is a \emph{non-blind} deconvolution method assuming knowledge of the ground-truth kernel; its results are provided for reference only.}
    \label{fig:bsd_qualitative}
\end{figure*}

\subsection{Quantitative Comparison}
In this section, we evaluate the proposed method \emph{Kernel-Diff} on synthetic and real blurred images and compare the performance with other state-of-the-art deconvolution methods. We evaluate our method on the \emph{BSD100} and \emph{RealBlur} datasets as described below along with the following deconvolution/deblurrring methods -- Deep Hierarchical Multi-Patch Network  (DHMPN) \cite{2019_DHMPN_CVPR}, Self-Deblur \cite{2020_SelfDeblur_CVPR}, Blind-DPS \cite{2023_BlindDPS_CVPR}, MPR-Net \cite{2021_MPRNet_CVPR}, and DeblurGANv2 \cite{2019_DeblurGANv2_ICCV}.  We use these methods as a representative set of contemporary deconvolution methods which include both iterative and end-to-end trained solutions. For the sake of a fair comparison, we retrain the end-to-end methods using synthetically blurred datasets with details provided in Section {\color{red}D} of the supplementary document.

\textbf{BSD100 Dataset}. We first evaluate our method on 100 synthetically blurred images from the BSD dataset \cite{2001_BSD_ICCV}. Specifically, we crop out random patches of size $256 \times 256$ and then blur them with synthetic motion blur kernels using symmetric boundary conditions. The performance of Kernel-Diff and other deblurring methods is shown in Table \ref{tab:bsd100} along with some qualitative examples in Figure \ref{fig:bsd_qualitative}.  We also compare estimated kernels from Self-Deblur, Blind-DPS and Kernel-Diff in Table \ref{tab:mnc} using maximum of normalized convolution (MNC) \cite{2012_RegionsToDeblur_ECCV}.
\begin{table*}[ht]
    \setlength\doublerulesep{1pt}
    \centering
    
    \begin{tabular}{c|c|c|c|c|c|c|g} 
        \toprule[1pt] \bottomrule[0.3pt]
        \makecell{Method $\rightarrow$ \\ Metric $\downarrow$} &  \makecell{DeblurGAN-v2 \\ \cite{2019_DeblurGANv2_ICCV}}& \makecell{Self-Deblur \\ \cite{2020_SelfDeblur_CVPR}}  & \makecell{Blind-DPS \\ \cite{2023_BlindDPS_CVPR}}& \makecell{DMPHN \\ \cite{2019_DHMPN_CVPR}}  & \makecell{MPR-Net \\ \cite{2021_MPRNet_CVPR} } & \makecell{\textbf{Kernel-Diff} \\ \textbf{(Ours)}} & \makecell{DWDN \\ \cite{2020_DeepWiener_NIPS}}  \\
        \hline
        PSNR $\uparrow$ &  15.44 & 13.53 & 17.56 & \underline{18.73} & 18.49 &  \textbf{18.95} & 22.95\\
        SSIM $\uparrow$  & 0.396 & 0.227 & 0.387 & 0.427 & \underline{0.435} &  \textbf{0.492} & 0.810\\
        LPIPS $\downarrow$ & \underline{0.563} & 0.665 & 0.670 & 0.617 & 0.632 & \textbf{0.335} & 0.169\\
        FID $\downarrow$ & 287.44 & \underline{259.40} & 280.53 & 265.05 & 265.60 &  \textbf{172.33} & 79.11\\
        \toprule[0.3pt] \bottomrule[1pt]
    \end{tabular}
    \vspace{0.0ex}
    \caption{\textbf{Performance on BSD100 dataset with synthetic blur}. \textbf{Bold} and \underline{underline} refer to overall and  second best blind deconvolution method respectively. DWDN is a non-blind method which knows the ground-truth kernel and is provided as an empirical upper bound only.}
    \vspace{-1ex}
    \label{tab:bsd100}
\end{table*}

\begin{table}[ht]
    \setlength\doublerulesep{1pt}
    \centering
    \begin{tabular}{c|ccc}
        \toprule[1pt] \bottomrule[0.3pt]
        Method & Self-Deblur & Blind-DPS & \makecell{\textbf{Kernel-Diff}}  \\ \hline
        MNC $\uparrow$ & 0.206 & 0.283 & \textbf{0.698}\\
        \toprule[0.3pt] \bottomrule[1pt]
    \end{tabular}
    \caption{\textbf{Kernel Evaluation}. Maximum of normalized correlation (MNC) for estimated kernels for iterative blind deconvolution methods.}
    \label{tab:mnc}
\end{table}

\textbf{RealBlur-50}.
To evaluate our blind deconvolution method on real blurred images, we use the the RealBlur-J dataset \cite{2020_RealBlur_ECCV}. However, since the focus of this paper is only spatially invariant blur and not other forms of blur, we crop random patches of size $256 \times 256$ from 50 blurred images in the dataset to reduce the possibility of violating the spatially invarying blur assumption. The results of this comparison are shown in Table \ref{tab:realblur}.  
\begin{table}[ht]
    \small
    \centering
    \setlength\doublerulesep{1pt}
    \begin{tabular}{ccccc}
        \toprule[1pt] \bottomrule[0.3pt]
        \makecell{  Metric $\rightarrow$ \\ Method $\downarrow$} & 
        \hspace{0.5ex} PSNR $\uparrow$ \hspace{-1ex} & 
        \hspace{0.5ex} SSIM $\uparrow$ \hspace{-1ex} & 
        \hspace{0.5ex} LPIPS $\downarrow$ \hspace{-1ex} & 
        \hspace{0.5ex} FID $\downarrow$ \hspace{-1ex} \\ \hline
        DeblurGAN-v2 & 14.98 & 0.583 & 0.289 & 129.06\\
        Self-Deblur & 16.15 & 0.483 & 0.459 & 172.32\\
        Blind-DPS & 19.82 & 0.607 & 0.320 & 175.95\\
        MPR-Net & \underline{23.63} & \underline{0.715} & \underline{0.213} & \textbf{113.11}\\
        \textbf{Kernel-Diff} & \textbf{25.21} &\textbf{ 0.779} & \textbf{0.187} & \underline{116.61}\\
        \toprule[0.3pt] \bottomrule[1pt]
    \end{tabular}
    \caption{\textbf{Performance on RealBlur-50}. \textbf{Bold} and \underline{underline} refer to overall and second best blind deconvolution method respectively.}
    \label{tab:realblur}
\end{table}

\subsection{Ablation Study}
\;\;\;\;\;\textbf{Effect of Non-Blind Solver Guidance} Each iteration of the diffusion consists of two sub-steps: (1) reverse diffusion using the trained dennoiser $\epsilon_{\theta}(\cdot)$ and (2) gradient descent on $\|\vy - \hat{\vk}_{0}\circledast\hat{\vx}_0 \|^2$ using the non-blind solver. In Table \ref{tab:ablation1}, we show the effect on performance of the second sub-step. Through the giant gap in performance with and without the guidance, we find that reverse diffusion is insufficient to converge to a good kernel estimate and hence, the experiment justifies the extra complexity required to guide the diffusion process to minimize the reblurring loss.

\begin{table}[ht]
    \centering
    \setlength\doublerulesep{1pt}
    \begin{tabular}{c|cccc}
        \toprule[1pt] \bottomrule[0.3pt]
        \makecell{Metric $\rightarrow$ \\ Method $\downarrow$} & PSNR $\uparrow$ & SSIM $\uparrow$ & LPIPS $\downarrow$ & FID $\downarrow$ \\
        \hline
        w/o DWDN   & 17.83 & 0.414 & 0.387 & 194.58 \\
        with DWDN  & \textbf{18.95} & \textbf{0.492} & \textbf{0.335} & \textbf{172.33}\\
        \toprule[0.3pt] \bottomrule[1pt]
    \end{tabular}
    
    \vspace*{1.0ex}
    
    \begin{tabular}{c c c}
        \includegraphics[width=0.30\linewidth]{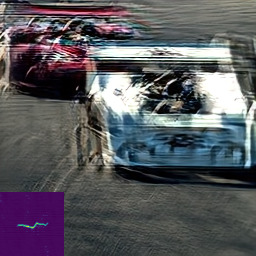} & \hspace{-2.0ex}\includegraphics[width=0.30\linewidth]{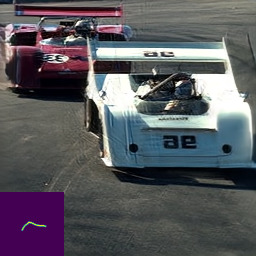} &
        \hspace{-2.0ex}\includegraphics[width=0.30\linewidth]{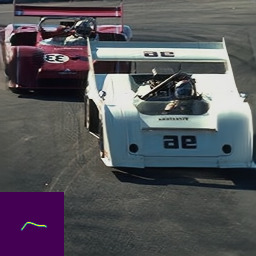}\\
        \makecell{\small{w/o non-blind} \\ \small{solver guidance}}  & \makecell{\small{with non-blind} \\ \small{solver guidance}}& \small{\makecell{DWDN \\ + True kernel}} 
    \end{tabular}
    \caption{\textbf{Effect of Non-Blind Solver Guidance}. Through this ablation study, we examine the effect of the second sub-step in the diffusion process which guides the diffusion process to minimize the reblurring loss using the non-blind solver DWDN \cite{2020_DeepWiener_NIPS}.}
    \label{tab:ablation1}
\end{table}

\begin{figure}[h]
    \vspace{-2ex} 
    \centering
    \includegraphics[width=0.95\linewidth]{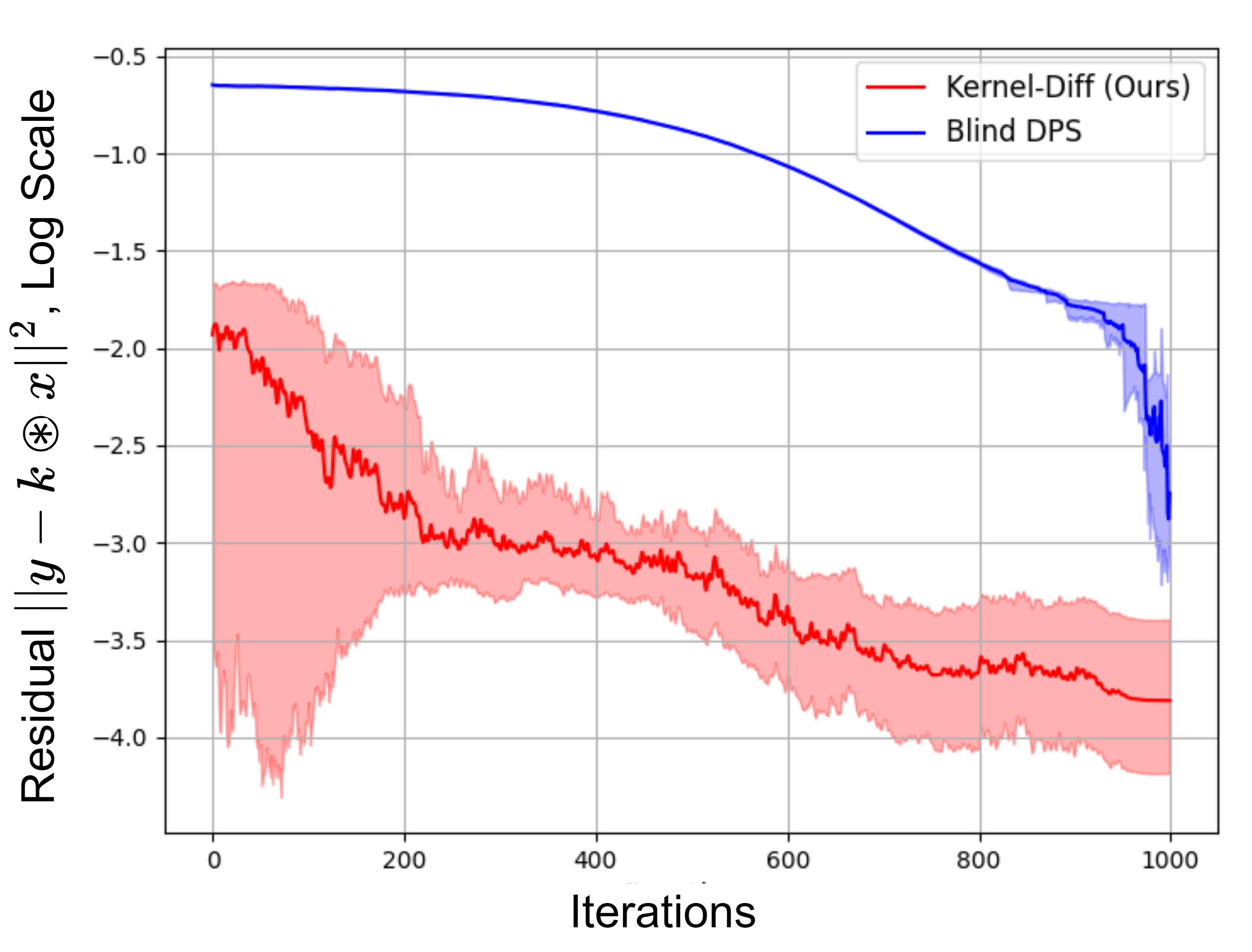}
    \caption{\textbf{Residual $\|\vy - \hat{\vk}_0\circledast \hat{\vx}_0\|^2$ for Blind-DPS and Kernel-Diff }. This experiments confirms our hypothesis that the proposed kernel estimation scheme converges to a better local minimum. The shaded region shows the range of the residual over 50 trials of each scheme. }
    \label{fig:reblur_loss}
\end{figure}

\textbf{Reblurring Loss during Diffusion} 
In Section \ref{subsec:motivation}, we show using a toy example why a alternate minimization approach is more likely to end up in a false local minima. We empirically demonstrate this by comparing the residual across the iterations, i.e., $\|\vy - \hat{\vk}_0 \circledast \hat{\vx}_0\|$ with Blind-DPS \cite{2023_BlindDPS_CVPR} in Figure \ref{fig:reblur_loss}. From the figure we observe that the residual in our proposed kernel estimation scheme converges to a solution with a much lower residual value. Assuming the kernel and image likelihoods to be approximately equal for both solutions, we conclude that our method leads to solutions with a higher log-likelihood, and hence, better quality results.

\section{Conclusion}
\label{sec:conclusion}
In this paper, we proposed Kernel-Diff, a blind deconvolution solution in the tradition of Levin et. al \cite{2011_LevinUnderstanding_PAMI}. We first show the pitfalls of conventional minimization using a 1D example and then show how the kernel minimization approach can be implemented in practice using a pre-trained non-blind solver. Using the principles outlined, we propose a guided diffusion method that samples from $p(\vk|\vy)$ to solve the blind deconvolution problem and show extensive numerical experiments on blurred images. 

For future work, the method can be refined to deblur images with spatially varying and motion blur i.e. degradation which do not fit the single-blur convolutional model. Alternately, the non-blind solver can be modified to be robust to inaccuracies in kernel estimation.

{
\small
\bibliographystyle{ieeenat_fullname}
\bibliography{references}
}
\clearpage
\input{supp}

\end{document}

%% file: preamble.tex
\usepackage[dvipsnames]{xcolor}
\newcommand{\red}[1]{{\color{red}#1}}

\input{commands.tex}
\usepackage{algorithm}
\usepackage{algpseudocode}
\usepackage{makecell}
\usepackage{mathtools}
\usepackage{pbox}
\usepackage{footmisc}
\usepackage{float}
\usepackage{comment}
\usepackage{multirow,multicol}
\usepackage{color,colortbl}
\definecolor{Gray}{gray}{0.9}
\usepackage[pagebackref,breaklinks,colorlinks,citecolor=cvprblue]{hyperref}
\definecolor{cvprblue}{rgb}{0.21,0.49,0.74}
\newcolumntype{g}{>{\columncolor{Gray}}c}

%% file: commands.tex
\usepackage{setspace}
\usepackage{graphicx}
\usepackage{multirow,multicol}
\usepackage[]{amsmath}
\usepackage{amsbsy}
\usepackage{amssymb}
\usepackage{amsfonts}
\usepackage{pifont}
\usepackage{balance}
\usepackage{fancyvrb}

{\begin{list}               
    {$\bullet$ \hfill}{
        \setlength{\leftmargin}{\parindent}
        \setlength{\parsep}{0.04\baselineskip}
        \setlength{\itemsep}{0.5\parsep}
        \setlength{\labelwidth}{\leftmargin}
        \setlength{\labelsep}{0em}}
    }
{\end{list}}

\providecommand{\cref}[1]{Chapter~\ref{#1}}

\providecommand{\fref}[1]{Figure~\ref{#1}}

\providecommand{\E}{\ensuremath{\mathbb{E}}}

\providecommand{\bydef}{\overset{\text{def}}{=}}

\providecommand{\mA}{\mathbf{A}}

\providecommand{\mI}{\mathbf{I}}

\providecommand{\mS}{\mathbf{S}}

\providecommand{\vk}{\mathbf{k}}

\providecommand{\vn}{\mathbf{n}}

\providecommand{\vt}{\mathbf{t}}

\providecommand{\vw}{\mathbf{w}}
\providecommand{\vx}{\mathbf{x}}
\providecommand{\vy}{\mathbf{y}}
\providecommand{\vz}{\mathbf{z}}

\providecommand{\red}[1]{\textcolor{red}{#1}}

%% file: supp.tex
\setcounter{page}{1}
\maketitlesupplementary

\section*{A. Toy Example - Implementation Details}
For the toy example proposed in Section 3.1 of the main document, we use the following parameters to generate the loss function plots. The vectors $\vx$, $\vk$, and $\vy$ are assumed to be of length $N = 128$. Given this length, we can assume the following bounds on image space parameters $0 \leq  a \leq  96$ and $0 \leq w \leq 32$. We limit the blur kernel width to $0 < \sigma \leq 3$. the ground-truth parameters for the experiment are set as $a_{true} = 64, w_{true} = 10, \sigma_{true} = 1.5$

For visualizing the 3-dimensional cost function in Figure 2, we use a modified version of the random projection technique \cite{2018_VisualizingLoss_NeurIPS} used for the visualization of loss functions in neural networks. Specifically, we take two random normal directions $\vz_1, \vz_2 \sim \mathcal{N}(0, \mI_3)$ corresponding to $\{a, w, \sigma\}$ space . Next we scale the two directions corresponding to the respective scales i.e. 
\begin{align}
    \vt_i = \Big[ 64\vz_i^{(1)}, 32\vz_i^{(2)}, \vz_i^{(3)} \Big]\;\;\; i = 1,2
\end{align}
Given the random choice of the directions $\vt_1$ and $\vt_2$, we need to make sure they are orthogonal, which we achieve by the following operation on $\vt_2$
\begin{align}
    \vt_2 \leftarrow \vt_2  - \frac{\vt_2^T \vt_1}{||\vt_1||^2}\vt_1
\end{align}
Next, we visualize the loss function along the random directions starting from the ground truth parameters $[a_{true}, w_{true}, \sigma_{true}]$ along the directions $\vt_1, \vt_2$.

\section*{B. Comparison with Levin et. al}
In this paper, we show why the kernel estimation strategy from \cite{2011_LevinUnderstanding_PAMI} is superior compared to alternating minimization followed by a diffusion-based method incorporating it. Despite the similarities, we come to slightly different conclusions from the original paper which are outlined here. 
\begin{enumerate}
    \item The toy example from \cite{2011_LevinUnderstanding_PAMI}, presented in Fig. 1 in the paper, depends on the choice of the sparse prior. However, one can argue that $\text{MAP}_{\vx, \vk}$ framework can be useful by a better choice of prior - which is easily possible given deep learning based solutions, which did not exist in 2011. Compared to that, the conclusions arrived in this paper do not depend on the choice of prior since we boil down the optimization problem to the smallest possible subspace i.e. $a, w, \sigma$.
    \item While \cite{2011_LevinUnderstanding_PAMI} arrives at the conclusion that that the $\text{MAP}_{\vx,\vk}$ strategy favours the no-blur solution, our conclusion is slightly different. Through our choice of the toy-problem, we find that the while the no-blur solution is not favoured, a gradient descent based scheme is still likely to get stuck in false minima, suggesting use of of the $\text{MAP}_{\vk}$ framework.
\end{enumerate}

\section*{C. Marginalization Approximation}
 In this section, we describe approximation in Eq. 15 from the main document in detail. We start with the conditional kernel distribution $p(\vk| \vy)$ as a marginalized distribution as follows:
 \begin{align}
     p(\vk|\vy) = \int p(\vx, \vk | \vy) d\vx 
    \label{eq:marginalize}
 \end{align}
 We approximate the integrand $p(\vx, \vk| \vy)$ using the Laplace approximation 
 \begin{align}
     p(\vx, \vk | \vy) \approx p(\hat{\vx}, \vk | \vy) \exp\Big[ -\frac{1}{2} (\vx-\hat{\vx})^T\mS^{-1}(\vx-\hat{\vx})\Big]
 \end{align}
 where 
 \begin{align}
     \hat{\vx} \bydef \underset{\vx}{\text{arg}\max}\Big[ p(\vx, \vk | \vy)\Big]
 \end{align}
 and 
 \begin{align}
     \hat{\mS} \bydef \Big[\frac{\partial^2}{\partial \vx^2} \log p(\vx, \vk | \vy)\Big]_{\vx = \hat{\vx}}
 \end{align}
 Applying this approximation to \eqref{eq:marginalize},
 \begin{align}
     p(\vk | \vy) &\approx \int p(\hat{\vx}, \vk |\vy) \exp\Big[-\frac{1}{2}(\vx-\hat{\vx})^T\mS^{-1}(\vx-\hat{\vx})\Big]d\vx
     \\
     &= p(\hat{\vx}, \vk | \vy) \int \exp\Big[-\frac{1}{2}(\vx-\hat{\vx})^T\mS^{-1}(\vx-\hat{\vx})\Big]d\vx \\ 
     &= p(\hat{\vx}, \vk | \vy) ( (2\pi)^n |\mS|)^{1/2}
 \end{align}
 here $n$ represents the dimension of the image space $\vx$. We further approximate $\mS$ to be independent of $\vx$, and hence a constant $C$, for sake of computational simplicity. However, future work can focus on providing a better approximation for $|S|$ and include it in the marginalization. 

 Next we examine the $\hat{\vx}$ and its relation to the non-blind solver.
 \begin{align}
     \hat{\vx} &\bydef \underset{\vx}{\text{arg}\max} \Big[ p(\vx, \vk  | \vy) \Big] \\
     &= \underset{\vx}{\text{arg}\max} \Big[ \frac{p(\vy | \vk, \vx) p(\vx)p(\vk)}{p(\vy)}\Big] \\ &= \underset{\vx}{\text{arg}\max} \Big[ \frac{p(\vy | \vk, \vx) p(\vx)}{p(\vy)}\Big] = \underset{\vx}{\text{arg}\max} \Big[ p(\vx|\vk, \vy) \Big]
 \end{align}
 Therefore, we have shown that $\hat{\vx}$ is the Maximum-A-Posteriori estimate for the non-blind problem problem i.e. estimating $\vx$ once the kernel $\vk$ is known.

\section*{D. Experimental Details}
We describe the configurations we use to train the methods for comparison as follows.

\textbf{DeblurGAN-v2 \cite{2019_DeblurGANv2_ICCV}.} We train the DeblurGAN-v2 using the official implementation available \footnote{ \url{https://github.com/VITA-Group/DeblurGANv2}} with the default configurations, parameters, and loss function. We use the same training scheme as stated in \cite{2023_BlindDPS_CVPR} for the majority. The backbone of the generator is a pretrained Inception-ResNet-v2, and we fine-tuned the generator 1.5 million iterations per epoch for 10 epochs. The batch size is 1.

\textbf{MPR-Net \cite{2021_MPRNet_CVPR}, DHMPN \cite{2019_DHMPN_CVPR}}: We adopt the official implementations available 
\footnote{ \url{https://github.com/swz30/MPRNet}, \url{https://github.com/HongguangZhang/DMPHN-cvpr19-master} }. We apply the default configurations and their proposed loss function. Again following \cite{2023_BlindDPS_CVPR}, we use the training scheme to train the model for 30k iterations with a batch size of 3. 

\section*{E. Kernel-Diff Hyperparameters}
\textbf{Diffusion Hyperparameters} For the reverse diffusion process, we use a total of $T = 1000$ steps and linear schedule for the noise variance $\{\beta_t\}$ as defined in \cite{2021_DifffusionUNet_NeurIPS}. For training the denoising network $\epsilon_{\theta}(\cdot)$, we use the $\ell_2$ norm, instead of the usual $\ell_1$ norm, because we empirically observe that training with the latter converges to kernels smaller than the target. This leads to ringing artifacts in the corresponding image output from the non-blind solver $F_{\phi}(\vy, \vk)$.

\textbf{Gradient Step Size}: For the reverse diffusion scheme, $\delta$ in Step 9 of Algorithm 1 is chosen similar to the heuristic from Blind-DPS i.e.
\begin{align}
    \delta \bydef \frac{0.1}{\|\vy - \hat{\vk_0}\circledast\hat{\vx}_0\|^2}
\end{align}
This ensure that the gradient descent becomes only aggressive in later stages of the diffusion when we're closer to the true solution. Conversely, gradient descent is relatively negligible in earlier stages of diffusion.

\section*{F. More Qualitative Results} 
\begin{figure*}[h]
    \begin{tabular}{ccccccc}
        \includegraphics[width=0.14\linewidth]{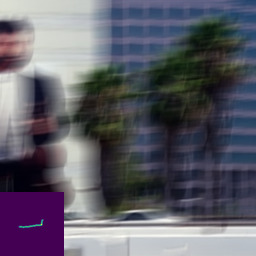} &
        \hspace{-2.0ex}\includegraphics[width=0.14\linewidth]{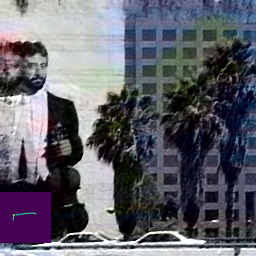} &
        \hspace{-2.0ex}\includegraphics[width=0.14\linewidth]{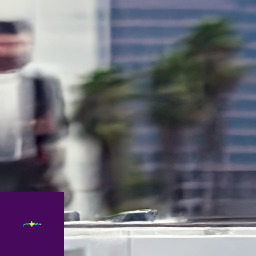} &
        \hspace{-2.0ex}\includegraphics[width=0.14\linewidth]{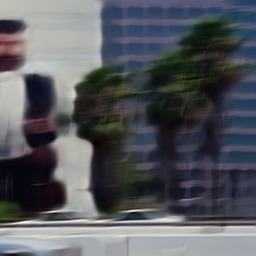} & 
        \hspace{-2.0ex}\includegraphics[width=0.14\linewidth]{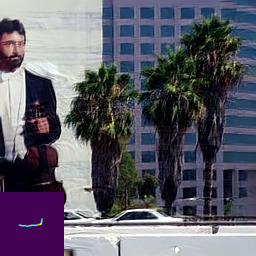} &
        \hspace{-2.0ex}\includegraphics[width=0.14\linewidth]{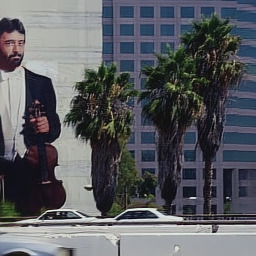} &
        \hspace{-2.0ex}\includegraphics[width=0.14\linewidth]{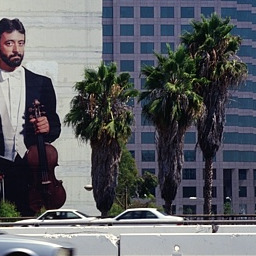}\\

        \includegraphics[width=0.14\linewidth]{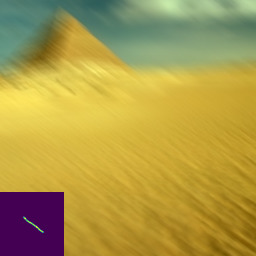} &
        \hspace{-2.0ex}\includegraphics[width=0.14\linewidth]{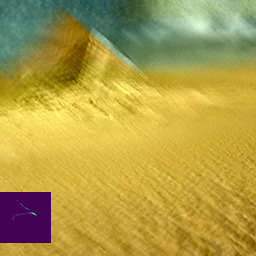} &
        \hspace{-2.0ex}\includegraphics[width=0.14\linewidth]{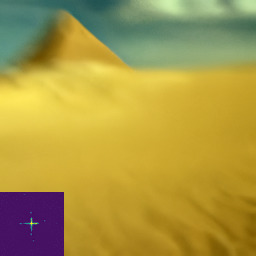} &
        \hspace{-2.0ex}\includegraphics[width=0.14\linewidth]{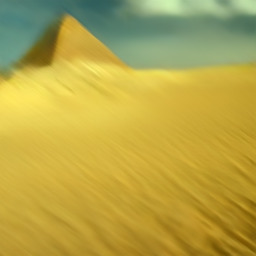} & 
        \hspace{-2.0ex}\includegraphics[width=0.14\linewidth]{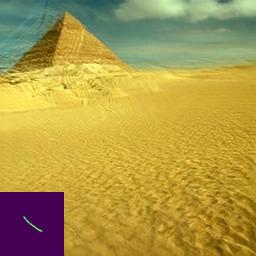} &
        \hspace{-2.0ex}\includegraphics[width=0.14\linewidth]{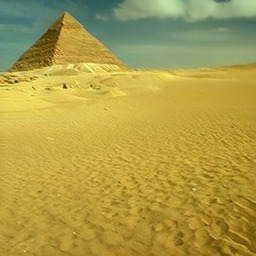} &
        \hspace{-2.0ex}\includegraphics[width=0.14\linewidth]{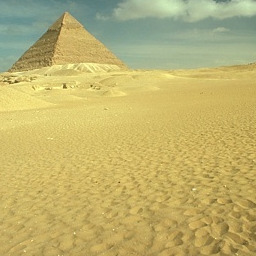}\\  
        \includegraphics[width=0.14\linewidth]{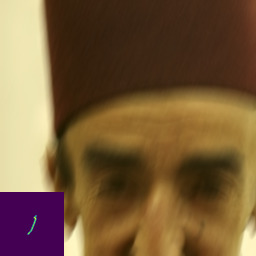} &
        \hspace{-2.0ex}\includegraphics[width=0.14\linewidth]{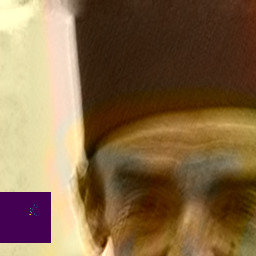} &
        \hspace{-2.0ex}\includegraphics[width=0.14\linewidth]{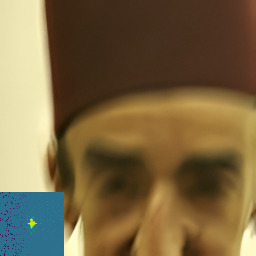} &
        \hspace{-2.0ex}\includegraphics[width=0.14\linewidth]{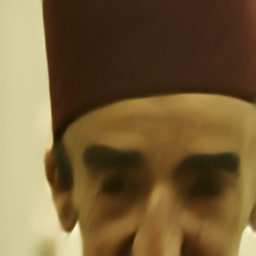} & 
        \hspace{-2.0ex}\includegraphics[width=0.14\linewidth]{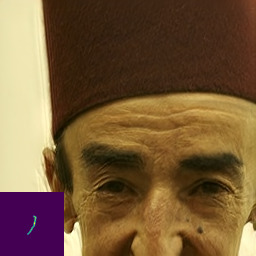} &
        \hspace{-2.0ex}\includegraphics[width=0.14\linewidth]{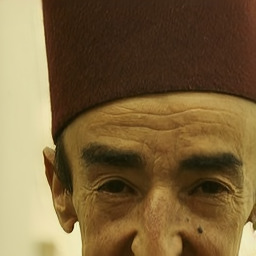} &
        \hspace{-2.0ex}\includegraphics[width=0.14\linewidth]{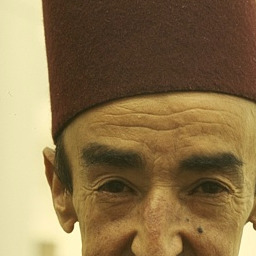}\\

        \includegraphics[width=0.14\linewidth]{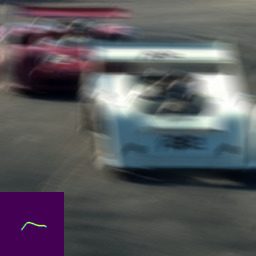} &
        \hspace{-2.0ex}\includegraphics[width=0.14\linewidth]{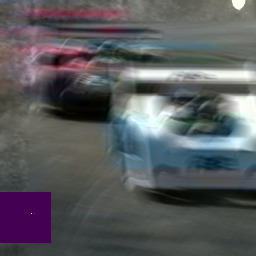} &
        \hspace{-2.0ex}\includegraphics[width=0.14\linewidth]{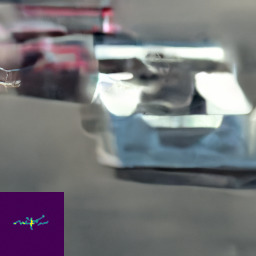} &
        \hspace{-2.0ex}\includegraphics[width=0.14\linewidth]{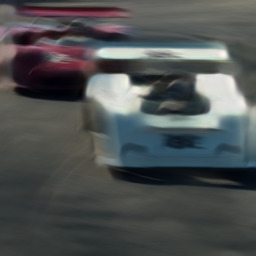} & 
        \hspace{-2.0ex}\includegraphics[width=0.14\linewidth]{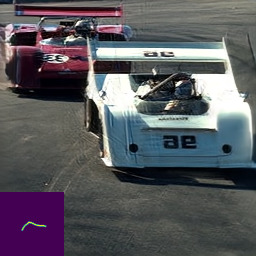} &
        \hspace{-2.0ex}\includegraphics[width=0.14\linewidth]{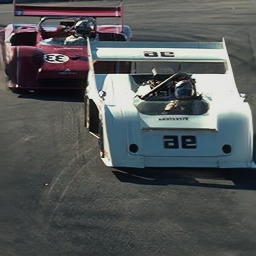} &
        \hspace{-2.0ex}\includegraphics[width=0.14\linewidth]{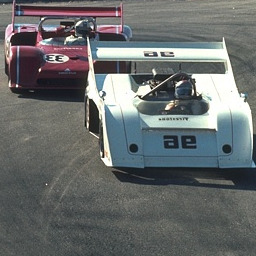}\\
        \includegraphics[width=0.14\linewidth]{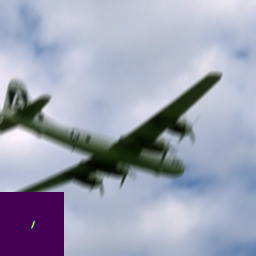} &
        \hspace{-2.0ex}\includegraphics[width=0.14\linewidth]{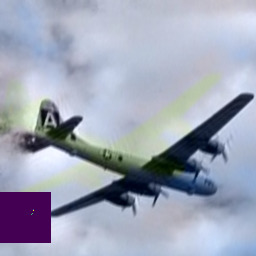} &
        \hspace{-2.0ex}\includegraphics[width=0.14\linewidth]{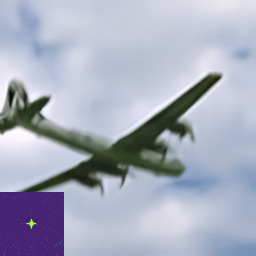} &
        \hspace{-2.0ex}\includegraphics[width=0.14\linewidth]{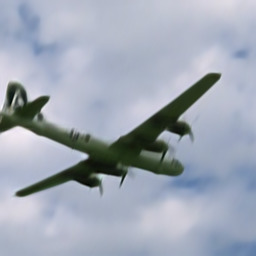} & 
        \hspace{-2.0ex}\includegraphics[width=0.14\linewidth]{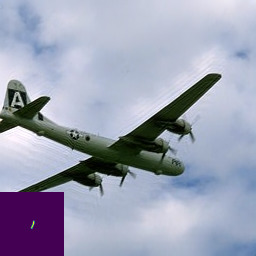} &
        \hspace{-2.0ex}\includegraphics[width=0.14\linewidth]{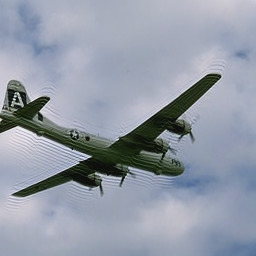} &
        \hspace{-2.0ex}\includegraphics[width=0.14\linewidth]{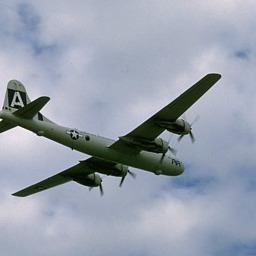}\\
        
        \includegraphics[width=0.14\linewidth]{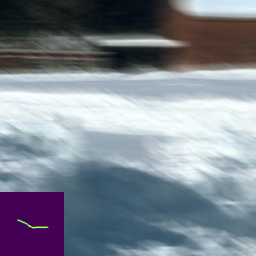} &
        \hspace{-2.0ex}\includegraphics[width=0.14\linewidth]{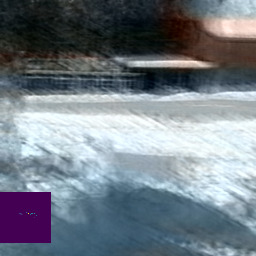} &
        \hspace{-2.0ex}\includegraphics[width=0.14\linewidth]{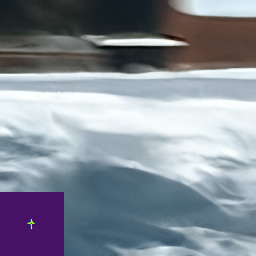} &
        \hspace{-2.0ex}\includegraphics[width=0.14\linewidth]{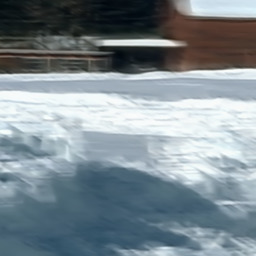} & 
        \hspace{-2.0ex}\includegraphics[width=0.14\linewidth]{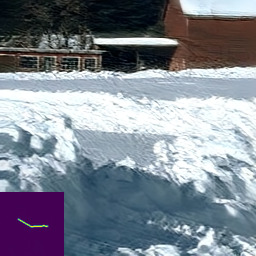} &
        \hspace{-2.0ex}\includegraphics[width=0.14\linewidth]{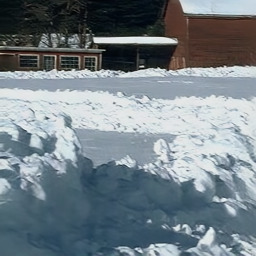} &
        \hspace{-2.0ex}\includegraphics[width=0.14\linewidth]{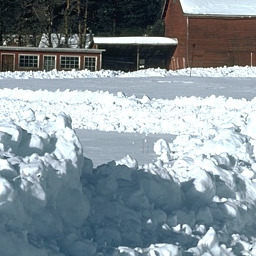}\\

        \includegraphics[width=0.14\linewidth]{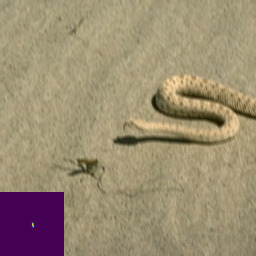} &
        \hspace{-2.0ex}\includegraphics[width=0.14\linewidth]{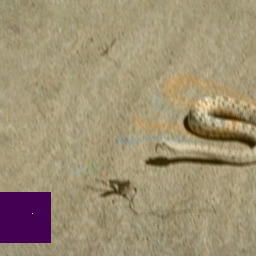} &
        \hspace{-2.0ex}\includegraphics[width=0.14\linewidth]{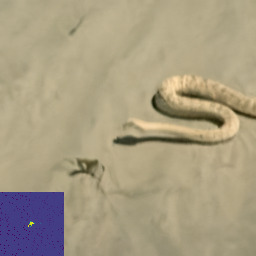} &
        \hspace{-2.0ex}\includegraphics[width=0.14\linewidth]{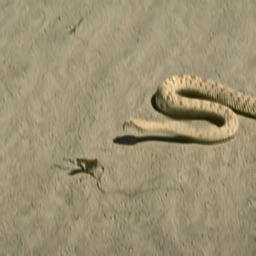} & 
        \hspace{-2.0ex}\includegraphics[width=0.14\linewidth]{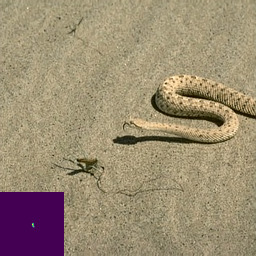} &
        \hspace{-2.0ex}\includegraphics[width=0.14\linewidth]{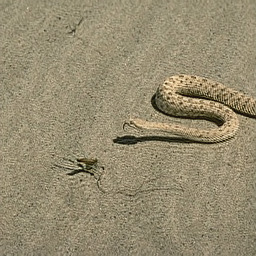} &
        \hspace{-2.0ex}\includegraphics[width=0.14\linewidth]{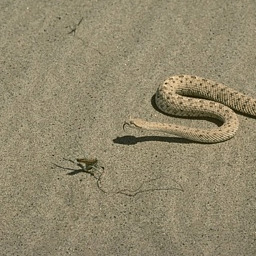} \\

        \includegraphics[width=0.14\linewidth]{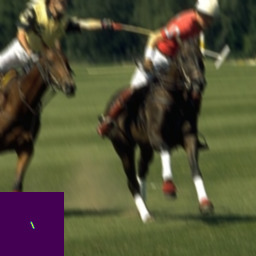} &
        \hspace{-2.0ex}\includegraphics[width=0.14\linewidth]{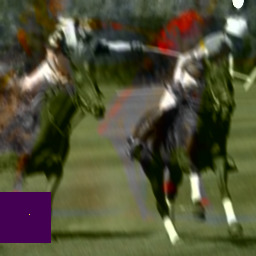} &
        \hspace{-2.0ex}\includegraphics[width=0.14\linewidth]{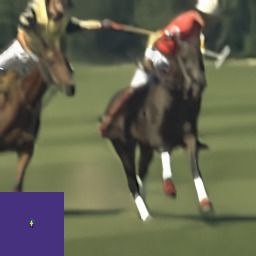} &
        \hspace{-2.0ex}\includegraphics[width=0.14\linewidth]{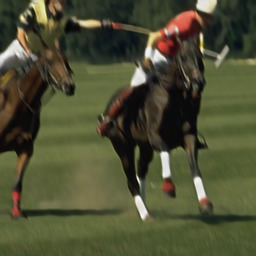} & 
        \hspace{-2.0ex}\includegraphics[width=0.14\linewidth]{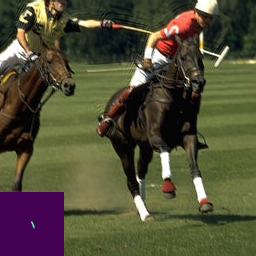} &
        \hspace{-2.0ex}\includegraphics[width=0.14\linewidth]{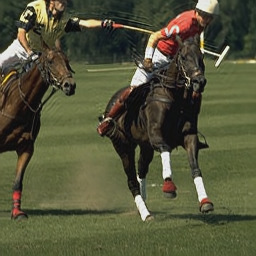} &
        \hspace{-2.0ex}\includegraphics[width=0.14\linewidth]{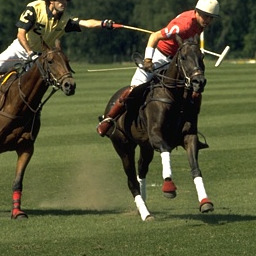}\\
        
        \makecell{\small{Blurred image} \\ \small{Inset: true kernel} } & 
        \hspace{-2.0ex}\makecell{\small{Self-Deblur} \\ \small{\cite{2020_SelfDeblur_CVPR}}} &
        \hspace{-2.0ex}\makecell{\small{Blind-DPS} \\ \small{\cite{2023_BlindDPS_CVPR}}} & 
        \hspace{-2.0ex}\makecell{\small{MPR-Net} \\ \small{\cite{2021_MPRNet_CVPR}}} & 
        \hspace{-2.0ex}\makecell{\small{\textbf{Kernel-Diff}} \\ \small{\textbf{(Ours)}}} & 
        \hspace{-2.0ex}\makecell{\small{DWDN-Oracle\textsuperscript{\red{\dag}}} \\  \small{\cite{2020_DeepWiener_NIPS}}} & 
        \hspace{-2.0ex}\makecell{\small{Ground Truth}} \\
    \end{tabular}
    \caption{\textbf{More Qualitative Results from BSD100}: Reconstruction results on synthetically blurred images from the BSD100 dataset.}
    \label{fig:bsd_qualitative_supp}
\end{figure*}

\begin{figure*}[h]
    \begin{tabular}{cccccc}
        \includegraphics[width=0.165\linewidth]{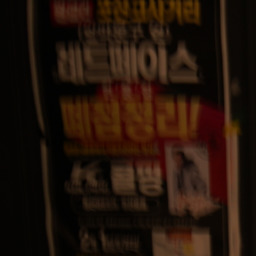} &
        \hspace{-2.0ex}\includegraphics[width=0.165\linewidth]{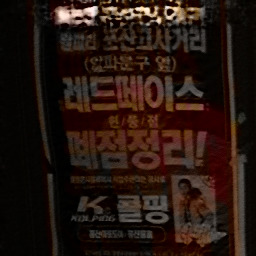} &
        \hspace{-2.0ex}\includegraphics[width=0.165\linewidth]{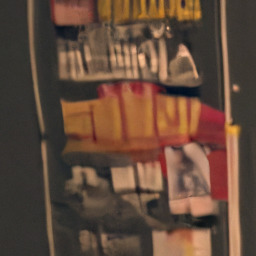} & 
        \hspace{-2.0ex}\includegraphics[width=0.165\linewidth]{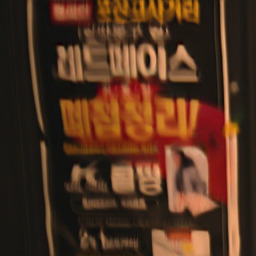} &
        \hspace{-2.0ex}\includegraphics[width=0.165\linewidth]{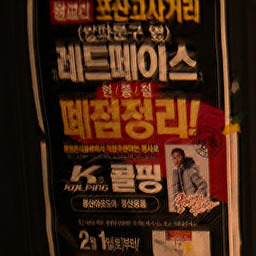} &
        \hspace{-2.0ex}\includegraphics[width=0.165\linewidth]{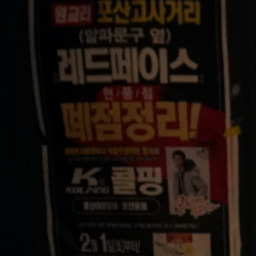}\\

        \includegraphics[width=0.165\linewidth]{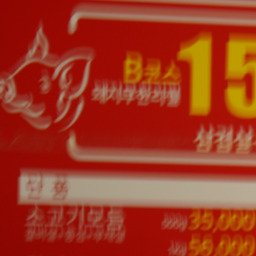} &
        \hspace{-2.0ex}\includegraphics[width=0.165\linewidth]{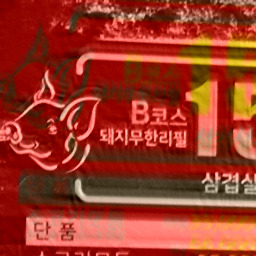} &
        \hspace{-2.0ex}\includegraphics[width=0.165\linewidth]{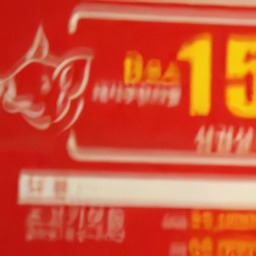} & 
        \hspace{-2.0ex}\includegraphics[width=0.165\linewidth]{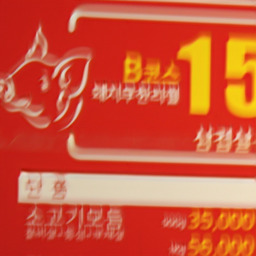} &
        \hspace{-2.0ex}\includegraphics[width=0.165\linewidth]{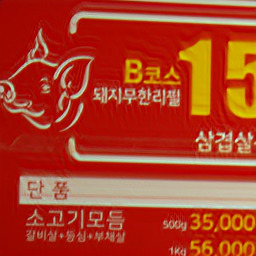} &
        \hspace{-2.0ex}\includegraphics[width=0.165\linewidth]{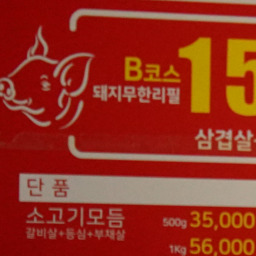}\\

        \includegraphics[width=0.165\linewidth]{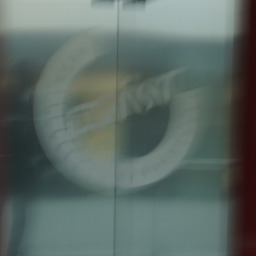} &
        \hspace{-2.0ex}\includegraphics[width=0.165\linewidth]{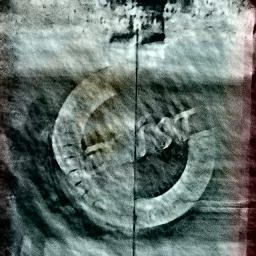} &
        \hspace{-2.0ex}\includegraphics[width=0.165\linewidth]{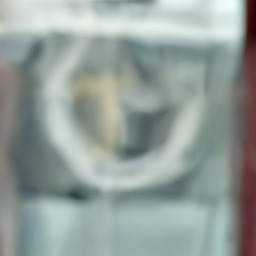} & 
        \hspace{-2.0ex}\includegraphics[width=0.165\linewidth]{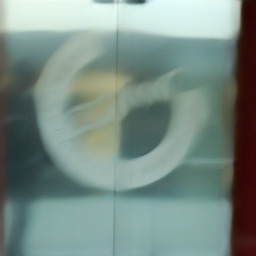} &
        \hspace{-2.0ex}\includegraphics[width=0.165\linewidth]{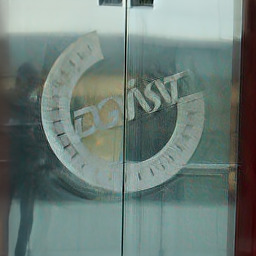} &
        \hspace{-2.0ex}\includegraphics[width=0.165\linewidth]{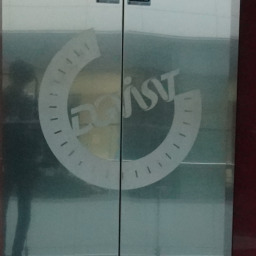}\\

        \includegraphics[width=0.165\linewidth]{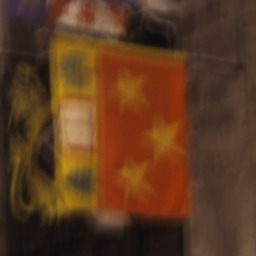} &
        \hspace{-2.0ex}\includegraphics[width=0.165\linewidth]{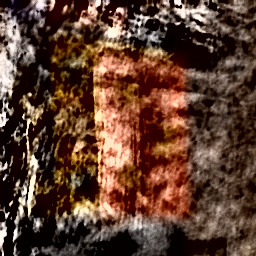} &
        \hspace{-2.0ex}\includegraphics[width=0.165\linewidth]{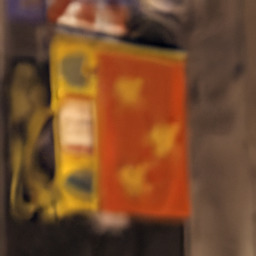} & 
        \hspace{-2.0ex}\includegraphics[width=0.165\linewidth]{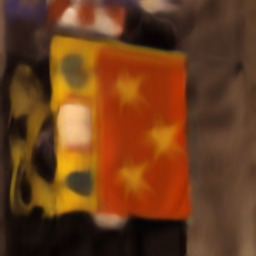} &
        \hspace{-2.0ex}\includegraphics[width=0.165\linewidth]{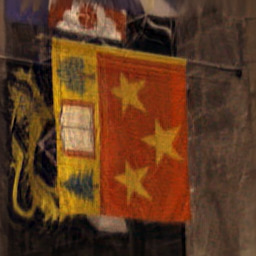} &
        \hspace{-2.0ex}\includegraphics[width=0.165\linewidth]{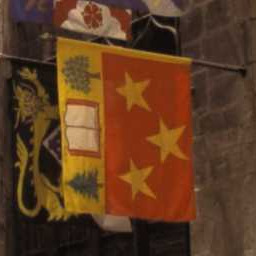}\\

        \makecell{\small{Blurred image} } & 
        \hspace{-2.0ex}\makecell{\small{Self-Deblur}  \small{\cite{2020_SelfDeblur_CVPR}}} &
        \hspace{-2.0ex}\makecell{\small{Blind-DPS}  \small{\cite{2023_BlindDPS_CVPR}}} & 
        \hspace{-2.0ex}\makecell{\small{MPR-Net} \small{\cite{2021_MPRNet_CVPR}}} & 
        \hspace{-2.0ex}\makecell{\small{\textbf{Kernel-Diff}}  \small{\textbf{(Ours)}}} & 
        \hspace{-2.0ex}\makecell{\small{Ground Truth}} \\
    \end{tabular}
    \caption{\textbf{Qualitative Results on Real Blurred Images} Comparison on real blurred images from the RealBlur \cite{2020_RealBlur_ECCV} and Kohler \cite{2012_Kohler_ECCV}. We would like to stress that real camera motion is \emph{not spatially invariant} due to in-plane rotation during exposure. \cite{2012_NUDeblurring_IJCV}. This may lead to artifacts while deblurring a single blur kernel. Note that even under this model mismatch, our method is able to obtain a relatively good blur kernel estimate  for deblurring.}
    \label{fig:realblur_qualitative_supp}
\end{figure*}

In Figure \ref{fig:bsd_qualitative_supp} and \ref{fig:realblur_qualitative_supp}, we provide more qualitative examples on synthetic and real blurred images reconstructed using our and contemporary methods. For the real blurred images in latter, we use $256 \times 256$ patches from the RealBlur \cite{2020_RealBlur_ECCV} and Kohler \cite{2012_Kohler_ECCV} dataset. 

For real-blurred images, we observe that the diffusion model as described in the main document is biased towards kernels generated in 
\footnote{\url{https://github.com/LeviBorodenko/motionblur}}. Therefore, to account for the domain gap, we retrain our method \emph{Kernel-Diff} using motion blur kernels generated from \cite{2011_BlurPerformance_TIP} and the corresponding results are shown in Figure \ref{fig:realblur_qualitative_supp}. Despite the retraining, results from Table 3 of the main document are provided using the original Kernel-Diff. Note that blur in images may not necessarily be spatially invariant due to in-plane camera motion during exposure \cite{2012_NUDeblurring_IJCV}. This may lead to artifacts caused by deblurring using a single estimated kernel.

